\documentclass[review]{elsarticle}

\usepackage{amsmath}
\usepackage{lineno,hyperref}
\usepackage{multirow}
\modulolinenumbers[5]
\usepackage{subcaption}

\begin{document}

\begin{frontmatter}

\title{Containment strategies and statistical measures for the control of Bovine Viral Diarrhea spread in livestock trade networks}



\author[TU,ERDC]{Jason Bassett\corref{Jason}}
\cortext[Jason]{Corresponding author}
\ead{jbassett.erdc@gmail.com}

\author[calimoto]{Pascal Blunk}

\author[FLI]{J\"orn Gethmann}

\author[FLI]{Franz J. Conraths}

\author[TU,UCC]{Philipp H\"ovel}

\address[TU]{Institut f\"ur Theoretische Physik, Technische Universit\"at Berlin, Hardenbergstrasse 36, 10623 Berlin, Germany}
\address[calimoto]{calimoto GmbH, Babelsberger Stra{\ss}e 12, 14473 Potsdam, Germany}
\address[FLI]{Institute of Epidemiology, Friedrich--Loeffler--Institut, S\"udufer 10, 17493 Greifswald - Insel Riems, Germany}
\address[UCC]{School of Mathematical Science, University College Cork, Cork T12 XF64, Ireland}
\address[ERDC]{U.S. Army Corps of Engineers Research and Development Center, Concord Massachusetts, U.S.A.}

\begin{abstract}
Assessing the risk of epidemic spread on networks
and developing strategies for its containment is of tremendous
practical importance, both due to direct effects in public health and its impact on economies. In this work we present the numerical results of a stochastic, event-driven, hierarchical agent-based model designed to reproduce the infectious dynamics of the cattle disease called Bovine
Viral Diarrhea (BVD), for which the corresponding movements' network is the main route of spreading. For the farm-node dynamics, the
model takes into account a vast number of breeding, infectious
and animal movement mechanisms via a susceptible-infected-recovered (SIR) type of dynamics with an additional permanently infectious class. The interaction between the farms is described by a supply and demand farm manager mechanism governing the network structure and dynamics. We discuss the disease and breeding dynamics, study numerous mitigation strategies of present and past government regulations taking Germany as a case-study country and perform a sensitivity analysis on key parameters. We argue that the model, subject to calibration, has universal predictive potential, can be extended to diseases beyond BVD and demonstrate that appropriate measures can indeed lead to eradication regarding BVD. We further present the time-series' results of the model and conduct a statistical analysis of and among the different mitigation strategies.
\end{abstract}


\end{frontmatter}


\section{Introduction}

Disease containment in the form of some intervention strategy is the main goal of any biosecurity program. In the last decades efforts on the behalf of governments have been implemented at country levels for the purpose of disease containment in livestock, especially for the European case, with the advent of integrated union policies. A major such disease that has been afflicting the livestock industry is \textit{Bovine Viral Diarrhea} (BVD) \cite{GRE03b,STA12,GET15,RIC2019}.

BVD is a disease caused by the virus of the pestivirus family. The main symptom of an ailing animal is diarrhea but general fatigue, fever and mucosal secretions are among the accompanying inflictions. Contagion can occur from all sorts of bodily fluids (including sneezing droplets) as well as from contaminated material or agricultural equipment. The symptoms persist for one to two weeks before the animal is recovered, provided complications, leading to its death, do not occur. The main peculiarity of BVD is its persistence when affecting early-stage embryos of carrying cows. If the embryos in question have not yet developed fully their immune system it is highly likely that they will carry the virus for the rest of their lives and be thus an enormous source of infection, provided they are born. Respectively, cows undergoing the disease are likely to be led to abortions \cite{LIN03b,LAN14}.

The impact of BVD on the agricultural industry, both beef and dairy, is immense due to losses in animals as well as due to the reduced milk yield. This has led to numerous studies exploring the impact of BVD, its containment practices and potential to the economy for different countries \cite{GRE03b,STA12,SAN15b,PIN17}. Such studies have been focused on the in-herd dynamics in an aggregated fashion regarding the animal population \cite{DAM15}, have explored the empirical network effect of cattle movements on BVD spread with simple dynamics \cite{TIN12a}, have combined the two aforementioned elements in metapopulation models generically \cite{COU10a} and in a multiscale, data-driven fashion (Italy) \cite{IOT17} for predicting the spatial spread of BVD investigating several intervention strategies, and have formulated expert systems to test present and potential policies based on data-driven simulations (Ireland) \cite{THU17}. Furthermore, there have been some instances of economic assessment of policies against BVD based on data \cite{MAR18a}, one based on the model presented in \cite{BAS18} and its results scrutinized in this paper \cite{GET19}.

Putting our model to the test on real data, we use the animal trade network of Germany as a testbed. The current regulations on the containment of BVD are based on the mandate issued over ten years ago and implemented in 2011. It has focused on the detection and removal of persistently infected (PI) animals by testing calves with an antigen (virus detection) test soon after their birth \cite{LEG08}. An amendment was introduced in 2016 envisioning a quarantine of 40 days on farms suspected of PI animals through testing \cite{LEG08a}. The results have been impressive in comparison to historical data, although they did not lead to eradication \cite{WER17}.

With the current work we aim to employ the refined approach of an agent-based model for the combined dynamics of animal movements and in-herd infections as opposed to aggregate methods \cite{BOS12}. This sort of modeling allows for a great level of detail to be included in the system. We additionally include a simulation plan of scenarios encompassing a wide range of intervention strategies emulating common biosecurity policies such as testing and vaccination. From that point we give an overview and interpretation of the resulting time-series for the global population as well as histograms for the infected animal populations at the farm level. Finally, we compare the statistical similarity of the time-series of the system running under no containment strategy with all the rest of the scenarios in pairs, utilizing the non-parametric Mann-Whitney U test.

\section{Materials and methods}


\paragraph{The model} 
We start with a short overview of the model that we developed. All the related details following the ODD (Overview, Design concepts and Details) protocol \cite{GRI10} can be found in a technical work of ours on the subject \cite{BAS18}.

For the simulation of BVD spread through the German cattle movement network we formulated an agent-based, stochastic, event-driven and hierarchical code in C++. We decided for the model to be agent-based (the animals being agents) with the aim to capture a great level of detail as reported in \cite{BOS12}. Furthermore, we introduced hierarchy in the sense that actions could correspond to different aggregates of agents (animals), thus enabling actions for a group of animals (herd), for a node in the network (the premises) and for the systems as a whole (for introducing intervention strategies). Stochasticity was something we also assumed necessary for both breeding and infectious features, so as to adequately model their complex, real world fluctuations. Moreover, we built the model in an event-driven fashion \cite{FIS13} in order to benefit from the trade-off between following the trajectory of an agent in continuous time and observing it only when a relevant, previously scheduled event is executed \cite{VES15a}.

\begin{figure}
	\centering
	\begin{subfigure}[b]{0.45\textwidth}
		\includegraphics[width=\textwidth]{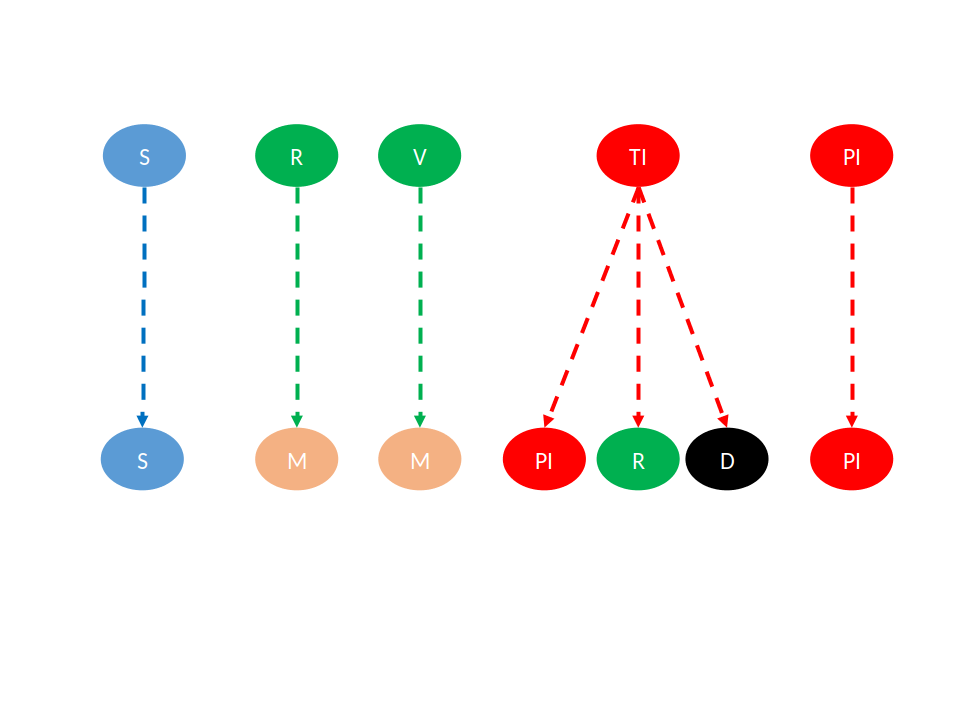}
		\caption{Vertical states.}
		\label{fig:SIRP_births}
	\end{subfigure}
	~
	\begin{subfigure}[b]{0.45\textwidth}
		\includegraphics[width=\textwidth]{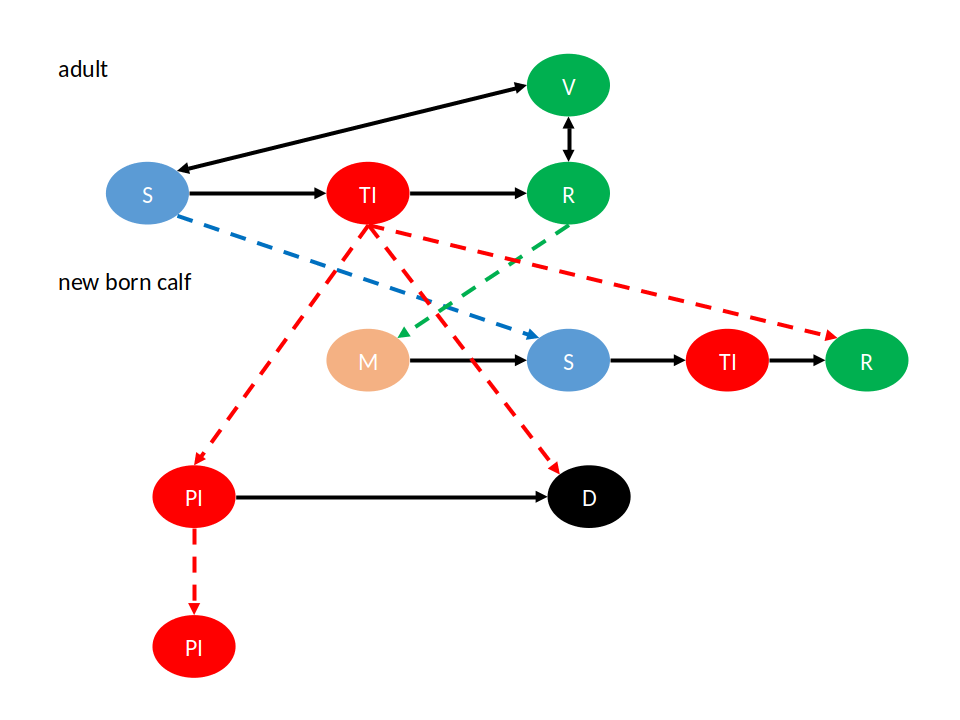}
		\caption{Combined horizontal and vertical states.}
		\label{fig:SIRP_full}
	\end{subfigure}
	\caption{Schematics (state charts) of the vertical (through births) \ref{fig:SIRP_births} and combined with horizontal (through contact) \ref{fig:SIRP_full} infectious states SIRP. The dashed lines denote vertical transmission paths (i.e. from mother to calf), while the solid horizontal ones. Note that a path to the demise of the animal is also taken into account with the letter ``D''. Additionally, the letters ``V'' and  ``M'' take into account temporary immunization effects induced by vaccination and protection through maternal antibodies respectively.}\label{fig:SIRP_schemes}
\end{figure}

Regarding the dynamics of the model, those were designated in two separate processes, namely to that of the in-farm (premise) dynamics and to that of the animal flow (movement) between the farms. For the former we employed the infectious SIR scheme with a permanently infectious class found in Viet et al. \cite{VIE04} (see figure \ref{fig:SIRP_schemes}), while for the latter a rule that every premise should demand or offer animals only when its population lacks or exceeds a certain quota respectively. In the case of the infectious, in-farm dynamics we assumed well-mixed conditions, i.e. no spatial structure and equal probabilities for all the susceptible animals to be infected. Accordingly, following standard SIR dynamics we defined a stochastic, instantaneous transmission rate for the infections in question, that dictated the inflow of infections in the pool of susceptible animals for each farm in a Markovian manner \cite{BAS18,VES15a}. For the latter dynamics, essentially the network dynamics, we supplied the farms, assuming solely dairy farms due to their high significance in the disease spread in Germany \cite{GET15,IOT17}, with the simplest supply and demand rule

\begin{equation}\label{eq:replacement}
\textrm{surplus/deficit} = \left| N_{\textrm{quota}} - N_{\textrm{instant}} \right|
\end{equation}

\noindent
i.e. the requirement that the number of animals offered or demanded (no special criterion applied for their selection) should be determined by the difference of an animal quota for each particular farm (specified upon initialization) $N_{\textrm{quota}}$ with the number of animals in the farm at the time of the movement $N_{\textrm{instant}}$. A surplus occurs when $N_{\textrm{quota}} > N_{\textrm{instant}}$ and a deficit in the opposite case. To moreover ascertain that the system is not dissipative nor inflationary (i.e. that relation \eqref{eq:replacement} for each farm is satisfied by the rest of the farms of the system), we include a source and a drain farm, which physically correspond to imports from outside the system (country) and slaughterhouses respectively. The precise details for both the infectious and the network dynamics can be found in our technical, ODD report \cite{BAS18}.

In addition, the model includes system-level intervention strategies including animal testing for their infectious status, non-targeted vaccination (immunization) as well as individual farm isolation (quarantine) for explicitly specified periods of time. Last but not least, we mention the sensitivity analysis on control parameters such as infectious transmission rate, vaccine and infectious identification (test) efficacy in \cite{BAS18}. The related code is publicly available on GitHub at the URL \url{https://github.com/Yperidis/bvd_agent_based_model} and follows the ``Overview, Design and Details'' protocol \cite{GRI10}.

\paragraph{Simulation setup}

For setting up and initializing the model's simulations we used expert opinion and the literature \cite{GET15,DAM15,TIN12a,IOT17,EZA07,HOS16,BIO16}. The exact details for the selection of all the parameters and the distributions employed can be found in the model's technical description \cite{BAS18}. The same holds for benchmarking with previous works of the literature.

\begin{figure}
	\centering
	\includegraphics[width=\linewidth]{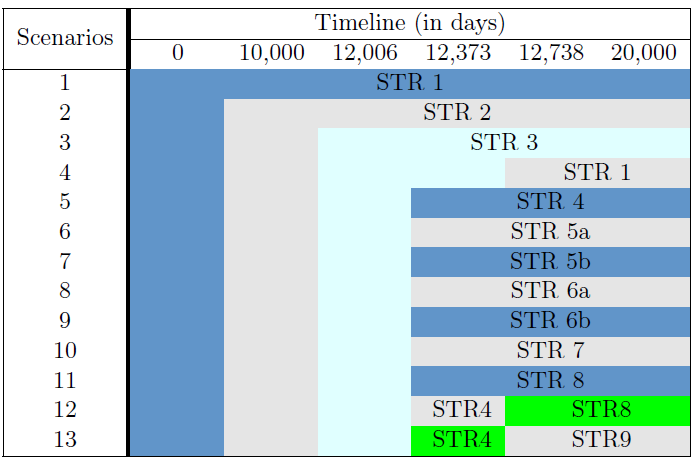}
	\caption{The scenario scheduling plan. STR stands for ``strategy'' as outlined in table \ref{tab:Strats}. A color block denotes the effect of the same strategy throughout the different starting days.}
	\label{fig:sim_setup}
\end{figure}

Here we only mention the minimal relevant settings, which can also be referenced in the simulation scheduling as summarized in tables \ref{tab:Strats} and \ref{fig:sim_setup} \cite{BAS19}.

\begin{itemize}
	\item The simulation was set to run for 20,000 time steps (the choice of the simulation parameters makes the time step a calendar day as explained in \cite{BAS18}).
	\item The units used for all the simulation's parameters mapped each time step to a calendar day. Thus the entire runs accounted to approximately 55 years.
	\item The involved tests' (antigen and antibody) sensitivity was set to 99.8\%, while their specificity was assumed to be 100\%.
	\item The probability that a vaccination's effect would successfully immunize the animal in question was set to 99.85\%.
	\item The infectious transmission coefficients, as defined in \cite{VIE04}, were set to $\beta_{\textrm{TI}}=0.03$ and $\beta_{\textrm{PI}}=0.5$.
	\item Only farms with populations over 10 animals have been included in the simulation. Smaller farms are deemed to have a negligent contribution to the spread of BVD in the network.
	\item The periodicity with which all the farms were queried as to their population status was set to 7 days.
\end{itemize}

For the terms \textit{old} and \textit{new regulation} as used in table \ref{tab:Strats}, a detailed explanation can be found in \cite{BAS19} and the original text of the regulations can be found at \cite{LEG08} and \cite{LEG08a}. \\


%

For the initialization of the farms' population we employed the farm size distribution of all of Germany scaled down to the size of the federal state of Thuringia. The scaling was performed due to computational resource limitations and we comment on that in the ``Discussion'' section. Regarding the initialization of the four different states of the population in each farm we first divided the farms into two classes: those having PI animals and those devoid of them. Then we set the first class to be a total of 2\% of all the farms. Furthermore, we set a 2\% probability for the animals entering the system in every instance of an import from the source farm to be PI, to account for a minimal BVD spread breaching the biosurveillance border system (this feature can be annulled and we make special reference to it in the supplementary material). Finally, we set the states of the animals for the population of each farm to be distributed according to the following probabilities for the two classes: (S, TI, R, PI)$_{\textrm{PI}}$=$(0.46,0.06,0.46,0.02)$ and (S, TI, R, PI)$_{\textrm{PI-free}}=(0.79,0.005,0.205,0)$, where S stands for ``susceptible'', TI for ``transiently infected'' and R for ``recovered'' (lifelong immunity). According to expert opinion from Dr. Gethmann of the Friedrich-Loeffler Institute, the aforementioned figures in a normalized population of farms and corresponding animals depicted the status of BVD in Germany upon the commencement of the nationwide biosecurity measures in 2011.

\begin{table}
	\begin{center}
		\begin{tabular}{|c|l|} \hline
			Strategy & Description \\ \hline
			1 		 & No control (baseline) \\
			2 		 & Old regulation \\
			3 		 & New regulation \\
			4 		 & New regulation and vaccination \\
			5a 		 & New regulation and YCW with a semesterly frequency \\
			5b 		 & New regulation and YCW with an annual frequency \\
			6a 		 & New regulation, vaccination and YCW with a semesterly frequency \\
			6b 		 & New regulation, vaccination and YCW with an annual frequency \\
			7 		 & YCW with a semesterly frequency \\
			8 		 & Vaccination \\
			9 		 & Vaccination and YCW with a semesterly frequency \\ \hline			
		\end{tabular}
		\caption{The different strategies comprising the scenarios presented in table \ref{fig:sim_setup}. YCW stands for the ``young calf window'' testing protocol.}\label{tab:Strats}
	\end{center}
\end{table}

\paragraph{Data sets}
As implied in the previous paragraph, the data-driven part of the model appeared upon the initialization with the input of a farm size distribution from a csv formatted file. This file consisted of two columns of integer numbers, one standing for farm number and the other for animal count. Each row linked the number of farms with the number of animals they corresponded to.

For the scaling of the German farm size distribution ($\approx 12^6$ animals in $\approx 157,000$ farms) to the size of the federal state of Thuringia ($\approx 350,000$ animals in $\approx 1,600$ farms) we simply multiplied all the entries of the farm column with a fraction of the total farm count for which the total animal count would be that of Thuringia ($\approx 350,000$). This number turned out to be $\approx 0.0279$.

\paragraph{Identification of infected animals}
The objective of any biosurveillance policy is to correctly detect infected animals. In the case of BVD the related policies focus on the detection of the PI animals either through an \textit{antigen} or an \textit{antibody} test \cite{LAN14}. Essentially the difference of the two consist of the direct detection of the virus, which is permanent in the PI animals, and the appearance of BVD virus antibodies in an animal indicating its infection (therefore transient) at some past point respectively.

Within the code we set two different modes of testing, accounting for both antigen and antibody tests.

\begin{itemize}
	\item \textit{Antigen test -- Regular blood test}
	This is a test that identifies infected (both transiently and persistently) animals. It is scheduled to be performed at the birth of an animal and is not repeated throughout the animal's lifetime. If an animal is identified as infected it is sent to the slaughterhouse within half a day. We envision additionally a challenge to a positive test outcome as a rare event, that leads to a second antigen test whose result determines irrevocable the fate of the animal in question. Due to the transient nature of the infection in otherwise healthy animals, the removal policy in question is targeted in its vast majority to the PI animals.
	\item \textit{Antibody test -- Young calf window (YCW) policy}
	This is a semesterly or annual test that identifies recovered animals similar to policies in Ireland and Scotland \cite{THU17,BRU10,TRA17}. It is performed at the specified times at the farm level, for a random selection of animals as summarized in table \ref{tab:YCW_table} with a methodology based on a hypergeomteric distribution sampling as outlined in \cite{CON15}. In case of a positive result an antigen test (regular blood test) is scheduled for every animal in the farm that has not been already tested. Note that, as in the antigen test case, in this manner only the PI animals are targeted.
\end{itemize}

	\begin{table}
	\begin{center}
		\begin{tabular}{|c|c|c|c|c|c|c|} \hline 
			N & $\leq$ 10 & $\leq$ 20 & $\leq$ 40 & $\leq$ 80 & $\leq$ 160 & $>$ 180 \\ \hline
			n & 8 & 10 & 12 & 13 & 13 & 14 \\ \hline 
		\end{tabular}
		\caption{Sample sizes given a farm population according to the young calf window method. Population sizes (N) and the corresponding samples (n) needed to be tested negative so that the prevalence of the infected animals in the population will not exceed 20\% with a confidence of 95\%.}\label{tab:YCW_table}
	\end{center}
\end{table}

%

\paragraph{Hypothesis testing -- Mann-Whitney U test}

Having any two sets of series of data with no specified assumption as to the form of their distribution (e.g. normal), one can use a non-parametric test to query about them following the same distribution (null hypothesis or $H_0$ valid) or not (alternative hypothesis or $H_1$ valid). Such a statistical test assuming \textit{independent} samples is the \textit{Mann-Whitney U test} \cite{mann1947,WAC14}.

The whole reasoning behind the Mann-Whitney U test lies in sorting the parts of the two independent (unpaired) samples in question in ascending order and measuring their absolute differences, i.e. the shift of their underlying distributions. The details of the implementation can be found in the original work \cite{mann1947} as well as in textbooks \cite{WAC14}. In our case any two time-series of the PI fraction of animals are the pairs of independent samples (because each one is initiated with a different seed in the stochastic simulation). Since our baseline scenario is No 1 we compare it pairwise to all the rest of the scenarios, i.e. we perform the test 12 times for the 13 scenarios. Finally, we set the significance level for the test to be 0.05 and run the wilcox.test function in the R language to produce a p-value. A p-value less than 0.05 will mean that the null hypothesis $H_0$ should be rejected otherwise it should be accepted.

\section{Results}

\paragraph{Disease extinction}

\begin{figure}
	\centering
	\includegraphics[width=\linewidth]{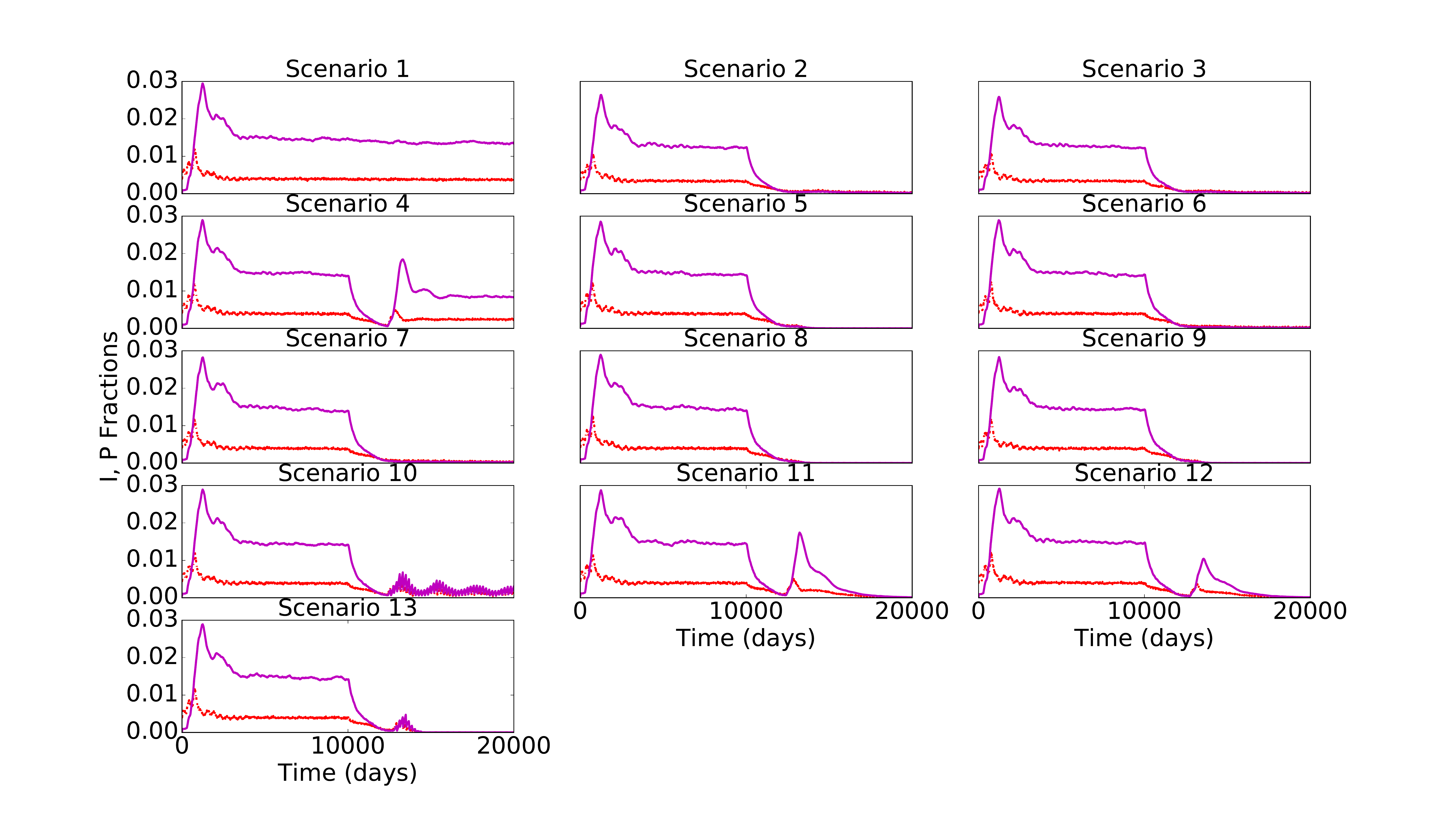}
	\caption{Transiently infected (dashed red line) and PI (dashed-dotted, magenta line) fractions of the population for scenarios 1 to 13.}
	\label{fig:IP_scen1_13}
\end{figure}

Having allowed the simulation to run as outlined in the plan \ref{fig:sim_setup} we obtained 13 different time-series for the four states of the system (susceptible, transiently infected, PI and recovered) as exhibited in figures \ref{fig:SIRP_scen1_13} and \ref{fig:IP_scen1_13}. Two questions that immediately arise are firstly, whether the PI class of animals is eradicated in any of the tested intervention strategies beyond a certain point in time, and secondly how fast (i.e. at which time) this eradication prevails.

A rather inconspicuous behavior that the system exhibited concerning the spread of BVD through the PI animals was their introduction through the source farm, taken that we had set a probability of 2\% for the incoming animals to be PI. All the animal introductions from the source farm for the given farm distribution with which the simulation was initialized ended at the 1,884\textsuperscript{th} time step (10,159 introductions in total) and respectively the last PI animal entering the system was observed at the 561\textsuperscript{st} time step. The former numerical figure demonstrates that for the given initialization the system quickly settles on a self-sustained trade to satisfy relation \eqref{eq:replacement} upon every management period (week), while the latter signifies the end of the BVD spread amplification from external triggering. To understand if the external PI amplification intensifies the epidemics we also ran the simulation for a 0\% probability of PI introductions upon animals entering the system through the source farm. In that case the introductions lasted until the 1,891\textsuperscript{st} time step (10,369 of total introductions), indicating again a state in time beyond which the system trade is self-sustained. In this latter case of no PI introductions in the system, the epidemic peak was slightly lowered compared to that of PI animals being introduced in the system with a 2\% probability (see the description of scenario 1 further on). Yet the PI prevalence remained the same in both cases by the 10,000\textsuperscript{th} time step, rendering the effect of the PI animals' introduction negligible for the scenarios of table \ref{fig:sim_setup}, which will be the focus of any further analysis. For a minimal example of how the PI prevalence is affected by PI introductions and their absence throughout the simulation reference the supplementary material (``Effect of PIs from the Source Farm'').

\begin{figure}
	\centering
	\includegraphics[width=\linewidth]{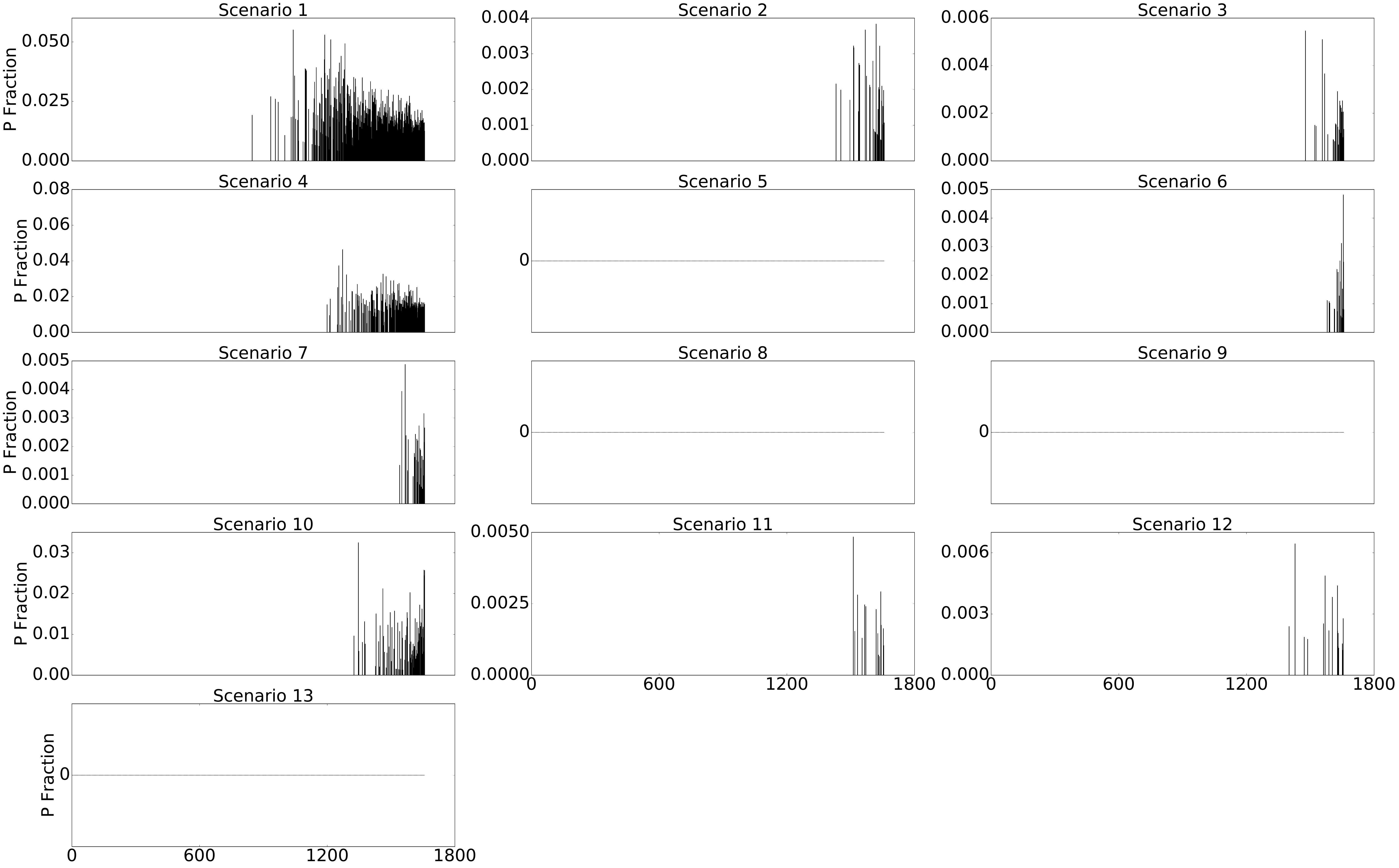}
	\caption{PI bar chart of the system's farms at the final time step for scenarios 1 to 13. Eradication is achieved for scenarios 5 (new regulation and vaccination), 8 (vaccination and YCW with a semesterly period), 9 (vaccination and YCW with an annual period) and 13 (new regulation, vaccination and YCW with a semesterly period). See table \ref{tab:Eradication_rankings}.}
	\label{fig:PI_distr_farms_scens}
\end{figure}

To address the question of PI eradication and its rate during the simulation, we draw the histograms of the PI population fraction for all the farms in ascending order of attributed ID numbers and for all the scenarios of plan \ref{fig:sim_setup}, as seen in figure \ref{fig:PI_distr_farms_scens}. Technically speaking, the ID attribution to the farms given from the farm size distribution is performed in a way such that the ascension of IDs corresponds to larger farms in terms of population. Thus, the aforementioned histogram also holds the answer to the farm size threshold above which there is a non-zero PI prevalence in scenarios where eradication is not achieved respectively.

The first direct result we notice by inspection of figure \ref{fig:PI_distr_farms_scens} is that PI eradication is achieved for scenarios 5, 8, 9 and 13 (see again tables \ref{tab:Strats} and \ref{fig:sim_setup}). What all these scenarios have in common throughout the duration of their intervention strategies (i.e. beyond the 10,000\textsuperscript{th} time step) is the periodical, non-targeted vaccination of animals combined with some removal through surveillance (testing) strategy, hinting that eradication cannot be achieved without both strategies.

Regarding the question of speed towards the eradication of the PI animals, scenario 8 holds the best records. This is intuitively expected as scenario 8 combines the new regulation with vaccination and the shortest (semesterly) antibody test scheduling per farm. Similarly, a little longer (annual) antibody test scheduling leads to the second best eradication time. What is not so obviously expected is the difference between scenario 13 and 5, which indicates that the YCW strategy when combined with vaccination (strategy 13) achieves faster PI eradication than its counterpart of the new regulation combined with vaccination (strategy 5). This is possibly a consequence of the stronger effect of the YCW semesterly testing when compared to the average animal arrival times, which in turns determines the time of the antigen test in the new regulation.  Note that the eradication time lags between the rankings translate to roughly a year.

	\begin{table}
	\begin{center}
		\begin{tabular}{cc|c}
			& & Eradication time \\	\hline		
			\multirow{4}{*}{Scenario ranking} 
			& 8 & 14,075 \\
			& 9 & 14,335 \\
			& 13 & 14,800 \\
			& 5 & 15,235 \\
		\end{tabular}
		\caption{Record times of PI eradication for the pertinent scenarios.}\label{tab:Eradication_rankings}
	\end{center}
\end{table}

Finally, the threshold farm size for PI animals to exist at the end of the simulation varies depending on the scenario we are looking at. In all the relevant scenarios, it is intuitive to expect the PI prevalence shifted towards the large farms of the input distribution, so that the SIR-sort of dynamics can work better (well-mixed conditions). The sole scenario however where the threshold is shifted to mid-size farms is the first, where the system has been left to evolve freely. This is the case where farms with a population of above 340 animals contribute to the persistence of the PI animals throughout the simulation runs.

\paragraph{Feature analysis of time-series}

As far as time-series are concerned, there are two prominent, distinct components we can discern. Firstly, the infectious states' evolution and secondly the population in absolute numbers. For the former we allow the discussion to revolve around the normalized population.

Regarding the global evolution of the infectious states in figures \ref{fig:SIRP_scen1_13}, \ref{fig:IP_scen1_13} (the latter focuses only on the infectious states) we should make a number of remarks for the interpretation of each scenario separately.

\begin{figure}
	\centering
	\includegraphics[width=\linewidth]{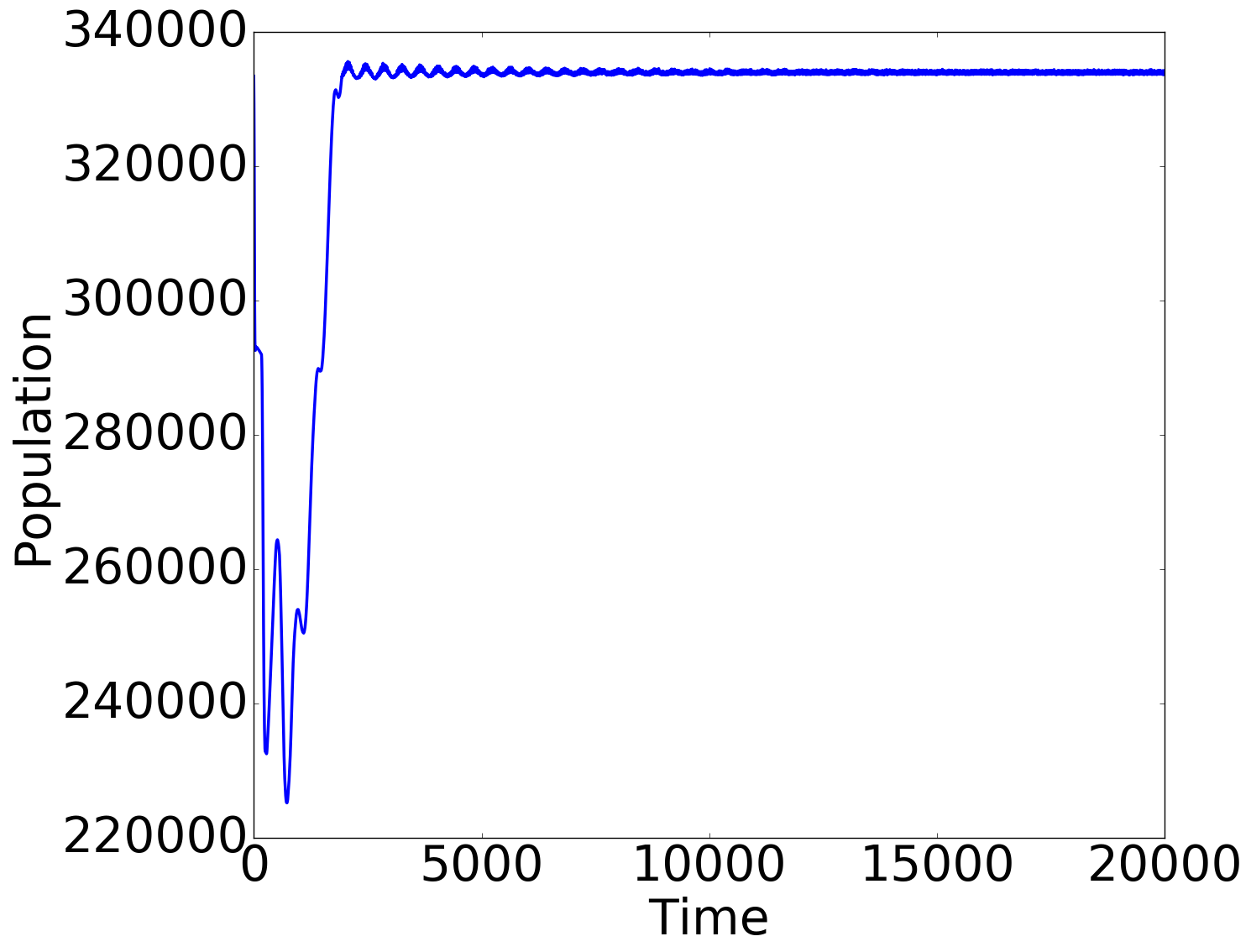}
	\caption{Global population evolution for the baseline scenario 1. The rest of the scenarios exhibit a statistically similar behaviour and are therefor not shown.}
	\label{fig:scena_base_pop}
\end{figure}

\begin{figure}
	\centering
	\includegraphics[width=\linewidth]{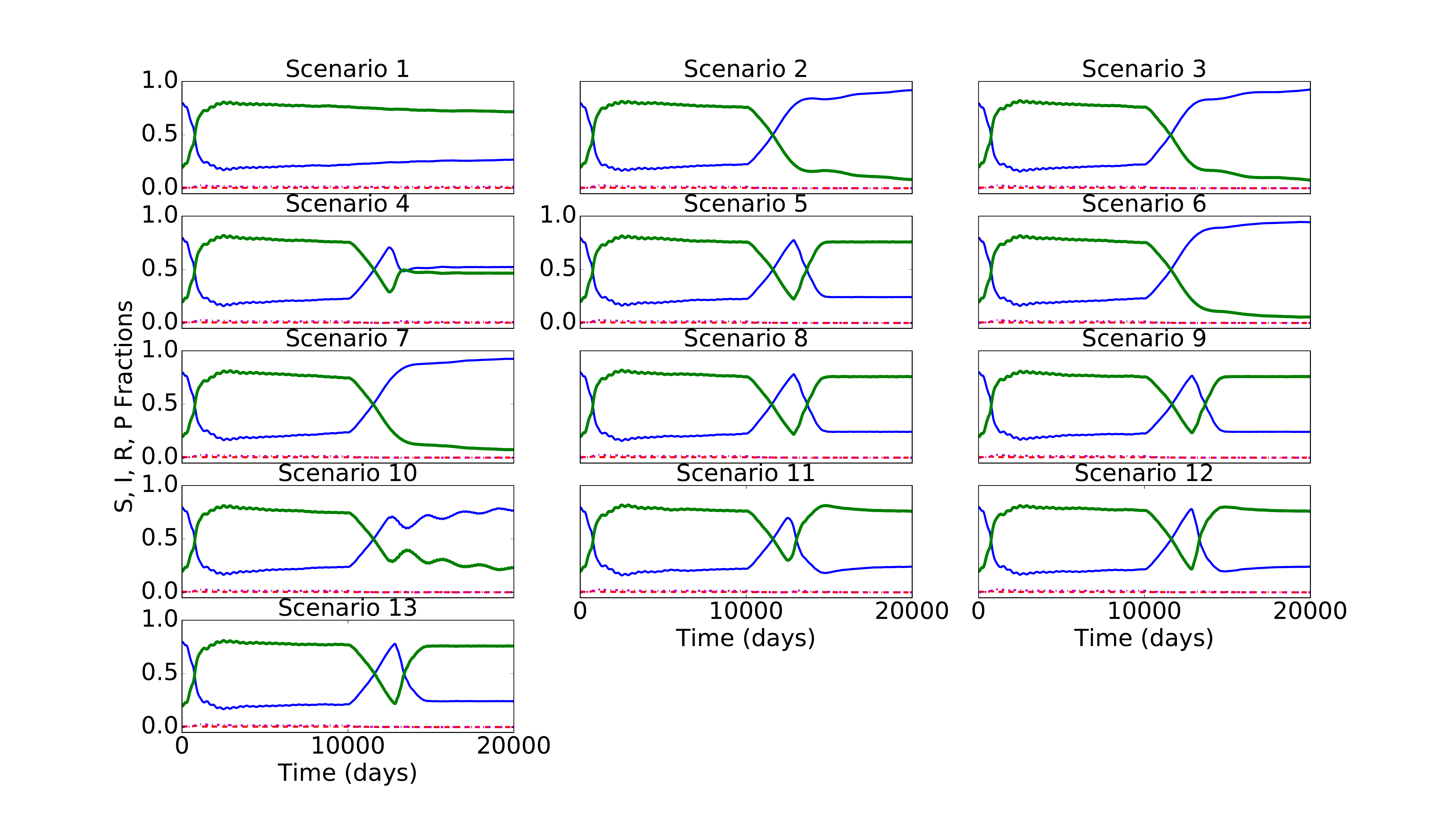}
	\caption{Susceptible (solid, blue line), transiently infected (dashed red line), recovered (dotted-dashed, green line) and PI (dashed-dotted, magenta line) fractions of the population for scenarios 1 to 13. Merely a magnification of figure \ref{fig:scena_base_pop} around the infectious states.}
	\label{fig:SIRP_scen1_13}
\end{figure}

\begin{itemize}
	\item \textit{Scenario 1 (baseline)}
	
	After a well-understood \cite{KEE05a,KEE08} initial epidemic transient peak ($\approx$ after 1,000 days) representing the outbreak from the initial infectious seeds within the farm level (see the initial conditions in the \textit{Materials and Methods} section) and from in-between farms (through movements) the system's infectious state fractions tend to an equilibrium state, with a PI prevalence of around 1.1\%. Correspondingly, the PI fraction distribution along all farms in figure \ref{fig:PI_distr_farms_scens} is the most pronounced of all the scenarios, which is logical as there is no counter measure. Up to the 10,000\textsuperscript{th} time step the behavior of all the scenarios is identical.
	
	\item \textit{Scenarios 2, 3}
	
	Beyond the 10,000\textsuperscript{th} time step the recovered (R) fraction drops significantly and the respective susceptible (S) increases. What happens is that due to the old regulations' effect the PI animals are being removed relatively fast (faster in the new regulation than in the old) and therefore the naive population, i.e. a population of S, is reinstated. When scenario 2 and 3 are compared graphically on the same scale there is no significant difference between them. However, the new regulation's effect in scenario 3 (partly a shorter time between two antigen tests and mostly the probation of a farm upon a PI detection) is more pronounced towards the eradication of PI animals. Taking into account the respective PI population fractions along the different farms in figure \ref{fig:PI_distr_farms_scens}, despite the fact that for scenario 3 we witness a higher PI fraction in some farms than in scenario 2, still the PI fraction distribution is narrower in scenario 3 than in scenario 2 supporting our previous assessment that the new regulation should be more effective than the old one towards the aim of the PI eradication. The higher PI prevalence in the farm distribution of figure \ref{fig:PI_distr_farms_scens} for scenario 3 is also a result of the input farm-size distribution's heterogeneities.
	Overall, scenario 3 depicts the current situation in Germany. The time frame between the 10,000\textsuperscript{th} day and the 12,006\textsuperscript{th} day corresponds roughly to the five and a half years (the half year difference with 2016 refers to a transition period from one mandate to the next) of the old regulation's effect in Germany.
	
	\item \textit{Scenario 4}
	
	Beyond the 10,000\textsuperscript{th} and up to the 12,373\textsuperscript{d} time step we observe as expected the same behavior as in scenario 3. After that point a behavior similar to that of scenario 1 reappears due to the lift of the intervention strategies. The final PI prevalence nevertheless appears to be slightly less than that of scenario 1, a fact which is reflected in the distribution width of the corresponding scenario of figure \ref{fig:PI_distr_farms_scens} when compared to the width of the distribution for scenario 1. This could be mainly attributed to the fact that at the point of the simulation where the intervention strategies are lifted the population has settled on a fixed point (see figure \ref{fig:scena_base_pop}), which is robust to the variations we implemented in the simulation plan (not presented due to minimal, if any, observed changes when compared to the baseline scenario in figure \ref{fig:scena_base_pop}). Thus the animals' movements could lead to fewer infections towards the steady state.
	
	\item \textit{Scenario 5}
	
	The behavior for this scenario is similar to scenario 3 with the essential difference starting from the 12,373\textsuperscript{d} step where vaccination is added to the new regulation. This action promotes the R population and conversely reduces the S leading the curves to cross for a third time in figure \ref{fig:SIRP_scen1_13} before they settle on a steady state. The fact that not all of the animals become immune at the steady state has to do with the interplay between the vaccination and the breeding dynamics, which sustain a non-zero susceptible population. As revealed in figure \ref{fig:PI_distr_farms_scens} this is one of the scenarios which leads to a PI extinction at the end of the simulation.
	
	\item \textit{Scenarios 6, 7}
	
	In scenarios 6 and 7 the effect of scenario 3 has nearly eradicated the PI animals therefore not exhibiting any dramatic difference from the YCW's implementation from the 12,373\textsuperscript{th} day onwards, either with a periodicity of a semester (scenario 6 with strategy 5a) or of a year (scenario 7 with strategy 5b). The two scenarios' behavior is similar to that of 3, with 6 (semesterly YCW testing) leading to a higher naive population reestablishment than 7 (annual YCW testing). This last observation is easier to see in the respective distribution of figure \ref{fig:PI_distr_farms_scens}: the distribution of scenario 6 is narrower than that of scenario 7.
	
	\item \textit{Scenarios 8, 9}
	
	Similarly to scenarios 6 and 7, in scenarios 8 and 9 the effect of scenario 3 has nearly depleted the PI population by the 12,006\textsuperscript{th} day to see any relevant effect afterwards. However, the inclusion of vaccination leads to a behavior similar to that of scenario 5 for that last segment of the simulation. The two scenarios, like 5, also lead to a final state PI eradication, albeit at faster rates than scenario 5.
	
	\item \textit{Scenario 10}
	
	In scenario 10 the YCW strategy beyond the 12,006\textsuperscript{th} day induces some periodicity in the PI population. This is an interesting effect which is attributed to the periodicity of the YCW test. The (declining) peaks' width of the PI population's fraction in figure \ref{fig:SIRP_scen1_13} signify the semesterly periodicity of the YCW test, which would remove PI animals generated in the time frame of no testing. This periodicity can also be seen by the declining ripples on the S and R curves in the corresponding plot of figure \ref{fig:SIRP_scen1_13}. This last effect would be induced by the infectious transmissions of the recurring PI animals. As suggested in figure \ref{fig:IP_scen1_13}, the respective distribution of figure \ref{fig:PI_distr_farms_scens} indicates that at the final state there is still some non negligent PI occurrence in the farms. Although in global population terms this PI prevalence seems to be rather low (less than 0.5\%) it is rather alarming that PI animals are relatively broadly distributed along the system's farms, introducing a risk of recurring future infections.
	
	\item \textit{Scenario 11}	
	
	Scenario 11 beyond the 12,373\textsuperscript{d} day is again similar in behavior to scenario 5 concerning the S and R curves in the corresponding plot of figure \ref{fig:SIRP_scen1_13}, although less pronounced than scenario 5. Looking closely at the respective plot of figure \ref{fig:IP_scen1_13} nonetheless we observe a short period where the PI population shoots upwards and is replenished to around the final state without strategy before dropping to minimal levels in the following segment of the simulation. This has to do with the success probabilities of the tests' accuracy and the working probabilities of the vaccination on the PI prevalence, namely that the indiscriminate vaccination scheduling protocol has a slower success rate than that of the ear tag testing. Especially for the peak observed after the 12,373\textsuperscript{d} day, it is a result of non-targeted vaccination (e.g. vaccination on recovered pregnant cows that are destined to produce a PI will not change the advent of the PI) in conjunction with the scheduling mechanism of the vaccination itself, which is intertwined with the insemination, as the latter needs to be scheduled at least 42 days after the vaccination. To demonstrate the inefficiency of non-targeted vaccination in one last manner, although the PI fraction has become very low, it is rather extraordinary that after 10,000 days (around 27 years) of the vaccination strategy's effect there is still a relatively (say to that of scenario 6) broad PI fraction distribution along the system's farms.
	
	\item \textit{Scenario 12}
	
	Scenario 12 combines the new regulation with indiscriminate vaccination	beyond the 12,373\textsuperscript{d} and up to the 12,738\textsuperscript{th} day following similar results to scenario 11. However, the post 12,738\textsuperscript{th} day behavior, which removes the new regulation's effect, exhibits again a similar, yet lower peak in PI as in scenario 11 before reaching minimal levels of the PI population. This can be attributed to a prolonged declining effect of PI-removed animals induced from the prior enforcement of the new regulation. Otherwise, in terms of the final state of the PI fraction's distribution along farms in figure \ref{fig:PI_distr_farms_scens} the situation is comparable to that of scenario 11 even with slightly broader and higher PI fractions. This demonstrates how unpredictably and inefficiently non-targeted vaccination can affect the PI prevalence of the system: even with a previously combined effect of vaccination with the new regulation, when vaccination is the sole intervention strategy it fails to lead to the eradication of the PI population even in a time-span of 26 years.
	
	\item \textit{Scenario 13}
	
	For scenario 13 the behavior is similar to scenario 12 up to the 12,738\textsuperscript{th} day. Beyond that we observe the first peak of the periodical PI increase (roughly extending to a semester again, which is the period of the YCW test) as in scenario 10, but with a smaller value due to the vaccination, which accompanies the YCW strategy at this stage. Henceforth the PI population is led to extinction by the vaccination's effect as can be seen in figure \ref{fig:PI_distr_farms_scens}, underlying once again the importance of testing in the enforced strategy towards PI eradication.
\end{itemize}

In respect to the population's evolution in absolute numbers (figure \ref{fig:scena_base_pop}), we have already commented on it through the infectious states. After the transient upon commencement due to the system's initial conditions, the population quickly settles on a fixed point, which is robust on the perturbations introduced by the different intervention strategies at various points in time. A striking feature that may be clearer after commenting is the decaying periodicity of the ripples observed in figure \ref{fig:scena_base_pop}. That is due to the initialized insemination events for the cows and their well-defined life-cycles, thus leading to the series of peaks, which are averaged collectively in time.

\begin{figure}
	\centering
	\includegraphics[width=\linewidth]{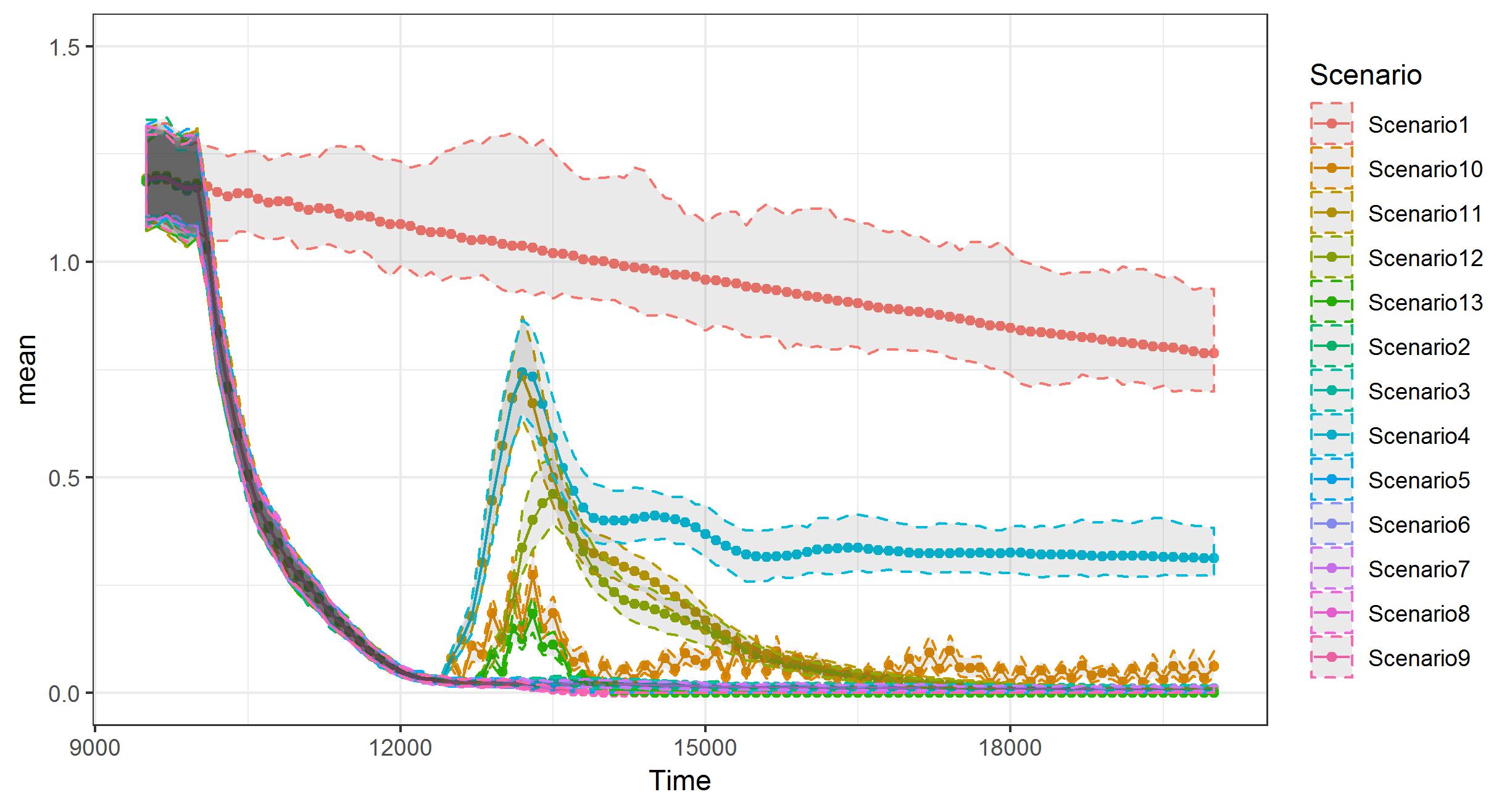}
	\caption{Variance around the mean PI prevalence ($\%$) for the 13 different scenarios at $t=13,000$.}
	\label{fig:MeanVar}
\end{figure}

\begin{figure}
	\centering
	\includegraphics[width=\linewidth]{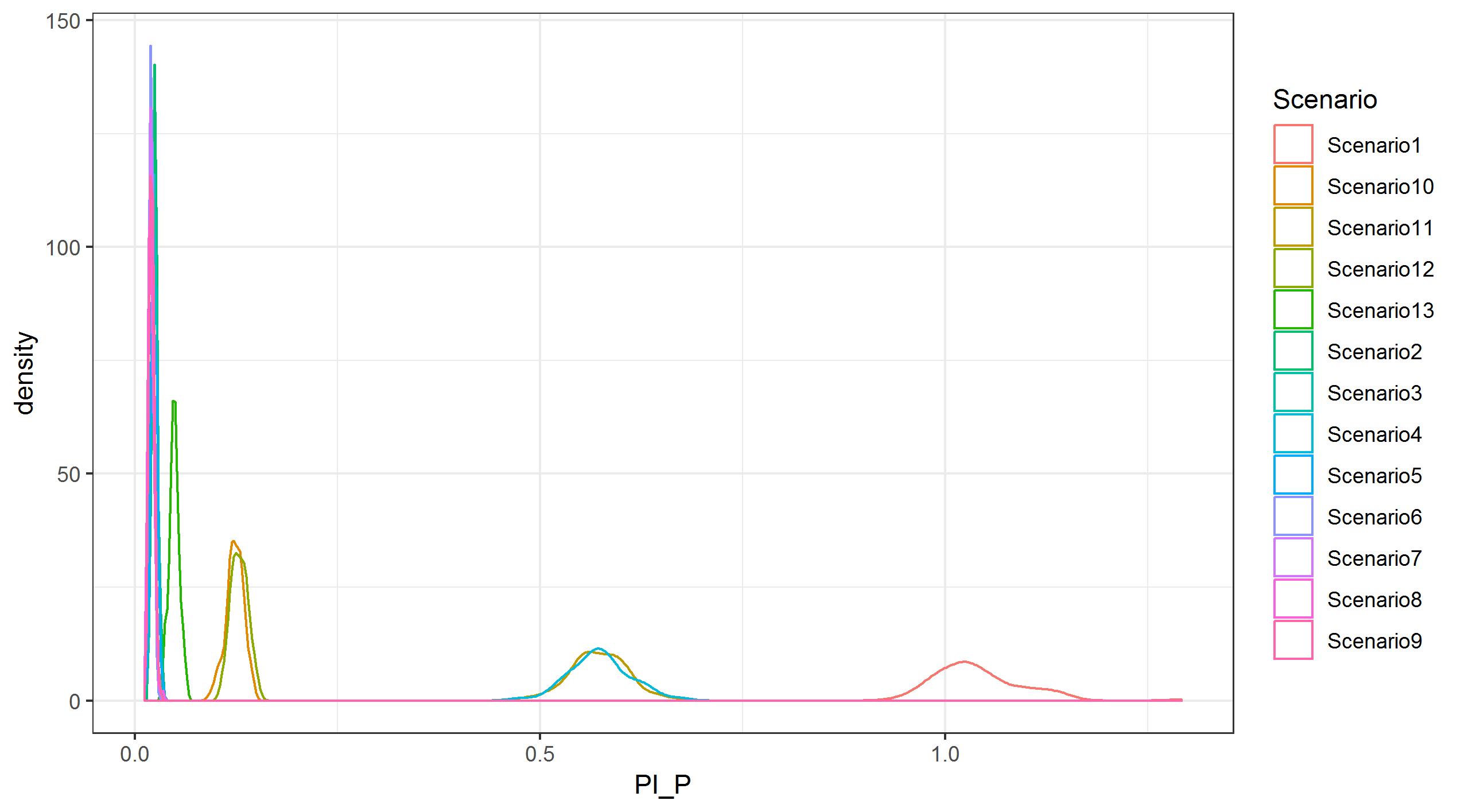}
	\caption{Variance distributions of the PI prevalence at t=13000 for the 13 different scenarios.}
	\label{fig:Var}
\end{figure}

Furthermore, in figures \ref{fig:MeanVar} and \ref{fig:Var} we explored the variance of the PI prevalence in the population (in gray) around a mean value and its distribution density at $t=13,000$ (i.e. the epidemic outbreak's peak exhibited in scenarios 4 and 10-13). The variance for each scenario was achieved by choosing 100 different integer seeds (ranging from 2,333,600,960 to 2,333,601,060) for the stochastic environment of the population and allowing it to run for each one of them. We performed those runs for all the 13 scenarios of the simulation plan \ref{fig:SIRP_scen1_13}. The fact that the PI population fraction mean is slightly dropping in time is not alarming as this is a mere indication that the fixed point is being approached at a rate so low that is insignificant for any practical purpose (i.e. policy making). The reason for this slow rate of decline in the PI population may be due to an underestimation of the transmission rates of BVD, which were taken from the literature \cite{VIE04}. Note that the times shown in figure \ref{fig:MeanVar} are only beyond the 9,000\textsuperscript{th} time step. That is because the strategies vary only after the 10,000\textsuperscript{th} time step. Extensive sensitivity analysis on the vaccination and testing can also be found in \cite{BAS18}.

\begin{figure}
	\centering
	\includegraphics[width=\linewidth]{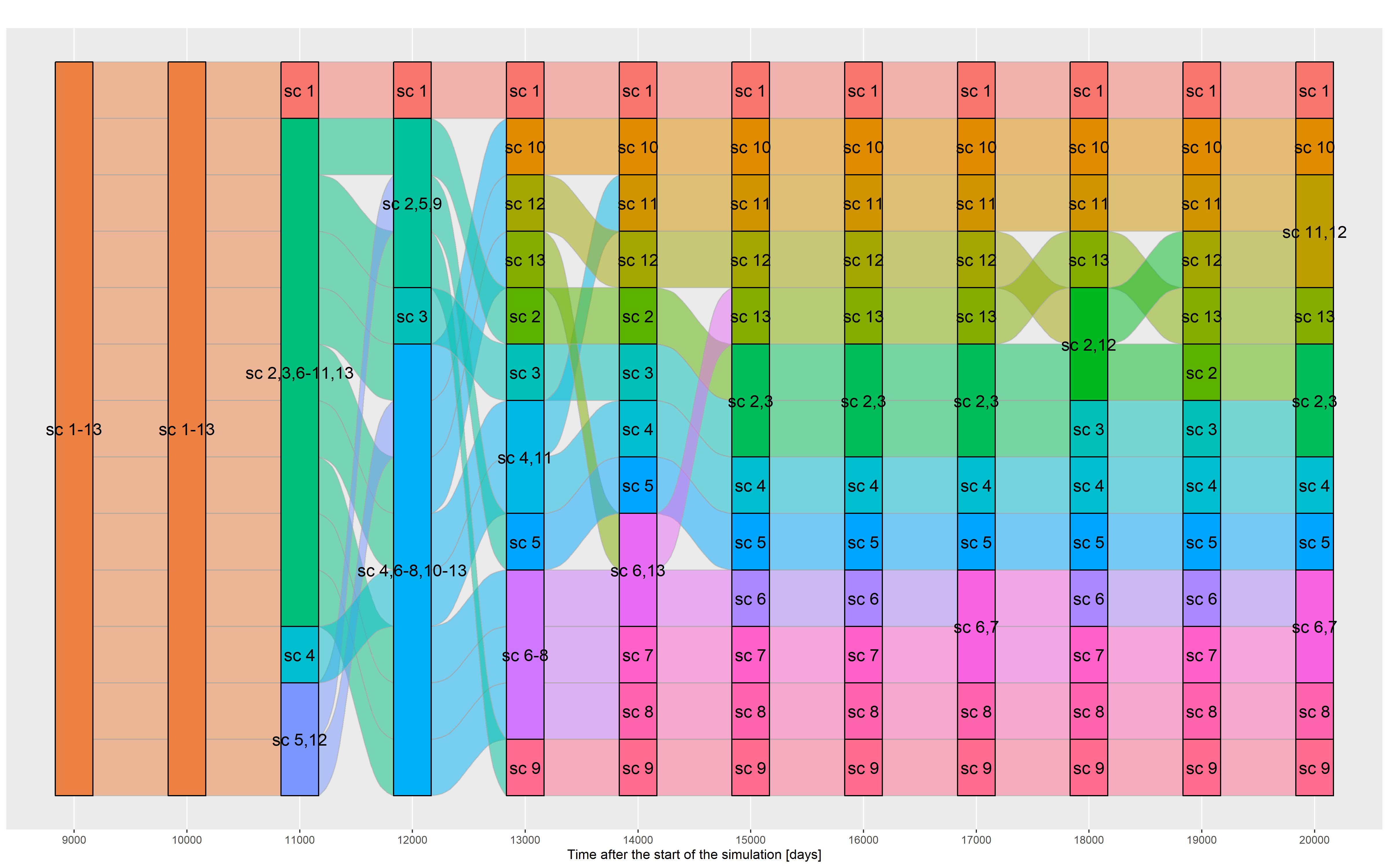}
	\caption{Similarity of the 13 scenarios of table \ref{fig:sim_setup} according to the Mann-Whitman U test.}
	\label{fig:classification}
\end{figure}

Finally, in figure \ref{fig:classification} we see the results of the Mann-Whitman U test for all the 13 scenarios in 13 blocks and grouped in different colors with a temporal increment (columns) of 1,000 time steps. Any changes start from the 10,000\textsuperscript{th} time step where strategies change according to table \ref{fig:sim_setup}. When the test accepts the null hypothesis for a number of scenarios, they are grouped with the same color within the column. Then for each increment the figure delineates the differentiation of the underlying population (i.e. the distribution of the infectious states) to the next for each block-scenario.

\section{Discussion}

Our results are among a series \cite{DAM15,TIN12a,THU17,VIE04} demonstrating the predictive power and insight that computational models can provide in the efforts of governance regarding epidemic spread containment in domesticated animal populations. For the case of BVD, we defined general scenarios with disease spread dynamics limited within the borders of a given country. These scenarios can be affected by control measures as well as the import of PI animals, which cannot be wholly prevented in practice when the laws of supply and demand are in effect and the country in question has no say over the biosurveillance regulations or their enforcement for the countries exporting to it. Significantly, all the results pinpoint to the importance of combined herd immunization (vaccination) and surveillance actions for achieving PI eradication, given a high sensitivity of BVD tests. As expected, when combined with vaccination, different surveillance practices lead to a range of record times up to eradication, with an antibody test augmentation of the current regulation topping all the alternatives. Furthermore, if vaccination is to be combined with biosurveillance programs then a refinement of the non-targeted practice can be certainly implemented with noticeable economical effects. In any case, the PI eradication times have to always be weighted against the foreseeable involved costs in the deployment of each scenario at the country scale. For the case-study of Germany such an analysis can be found in \cite{GET19}. Finally, for an optimal policy implementation the cost reduction induced by antibody test practices (YCW strategy), namely the bulk milk tank or serum test pool versus the individual blood tests should be taken into account.



Regarding the establishment of BVD in a farm population our results hinted on a critical population size of the farms leading to a persistent epidemic without intervention. In the cases where intervention still led to an endemic final state the farm sizes contributing to the epidemics were shifted to the largest farms of the systems. This is a consequence of the SIR dynamics taking place at the farm level: a well-mixed, large population with endemic conditions (infectious versus recovery rate ratio) including demographic changes can exhibit an endemic equilibrium point \cite{KEE08}. Nevertheless, the simulation being stochastic, one can expect fluctuations of the farm size threshold for endemicity.

In terms of extensions of the current work, the influence of the scaling on the predictions of the system should be investigated. Computationally the inclusion of more farms requires exponentially more resources in memory even for small farms, while increasing the number of animals in very few farms depletes memory resources much slower. Scaling the system to less farms and population would give a better understanding of the finite-size effects (lower bound) on the epidemic dynamics. Another parameter that can be further inquired is the sensitivity of the infection spread among farms to the farm size distribution given initially, i.e. to the heterogeneity of the distribution. Moreover, assuming the cattle movement rules are reasonably similar in most E.U. countries due to unified mandates and similar practices, the code employed in this work could be applied to extract results for at least any country within the union. Additionally, the inclusion of subtleties such as weighted or biased trading or dynamics at the node level would delve to an ever greater level of detail, attempting to capture features such as the preferential trade (attachment) or the observed declining trend of farms in Germany in the last decade (according to the federal statistical bureau of Germany) respectively. Last but not least and from a theoretical point of view, the computation of a threshold quantity from parameters of the simulation for the epidemic outbreak could be computed in a metapopulation manner as in \cite{COL08}. From there onwards various vaccination strategies for herd immunization can be encompassed to alter the epidemic thresholds as in \cite{KEE11}.

\section*{Acknowledgments}

J. Bassett and P. H\"ovel acknowledge the support of the German research association within the framework of the collaborative research center 910.







\section*{Supplementary Material}
\subsection*{Single Farm Dynamics and PI Effect from Source Farm}

\subsubsection*{Single Farm Dynamics}

\begin{figure}
	\centering
	\begin{subfigure}[b]{0.45\textwidth}
		\includegraphics[width=\textwidth]{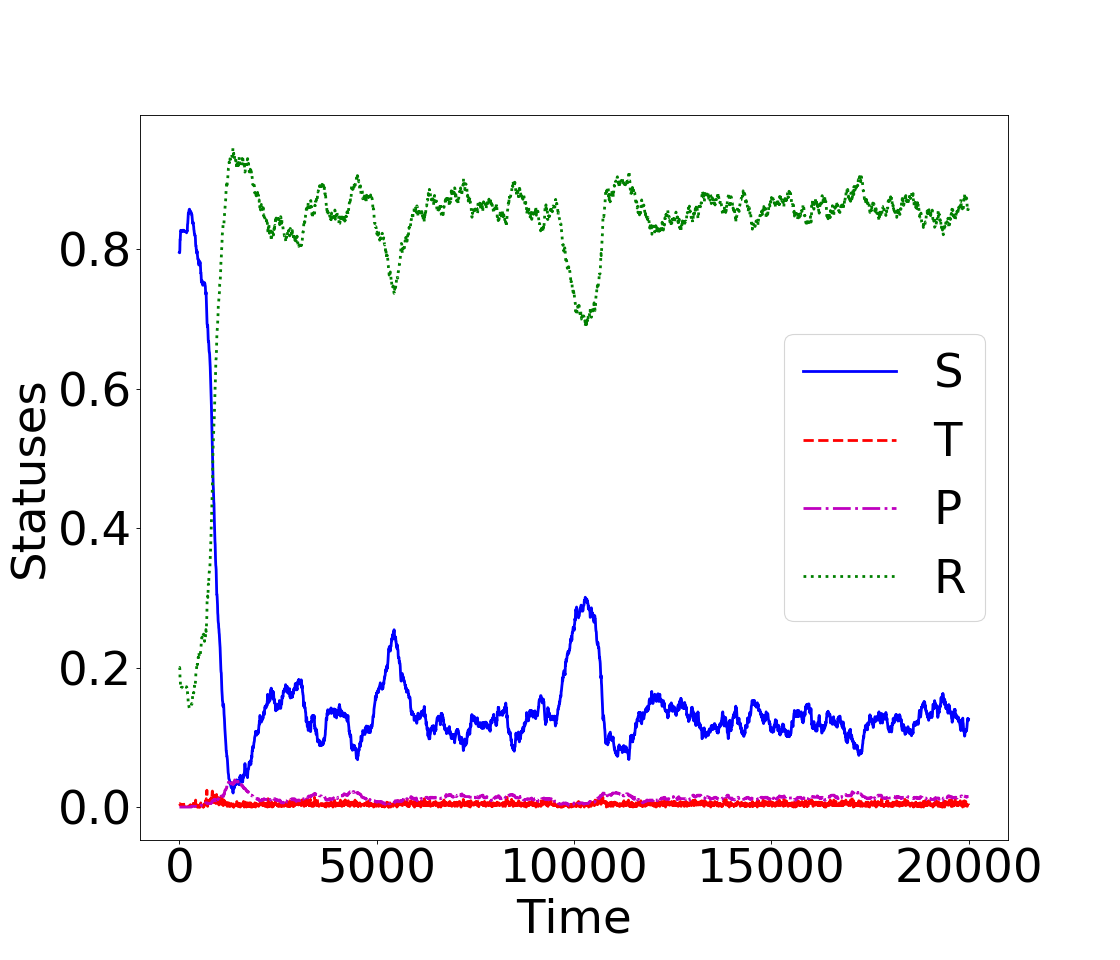}
		\caption{All states' evolution.}
		\label{fig:1farmSIRP_wellP0}
	\end{subfigure}
	~
	\begin{subfigure}[b]{0.45\textwidth}
		\includegraphics[width=\textwidth]{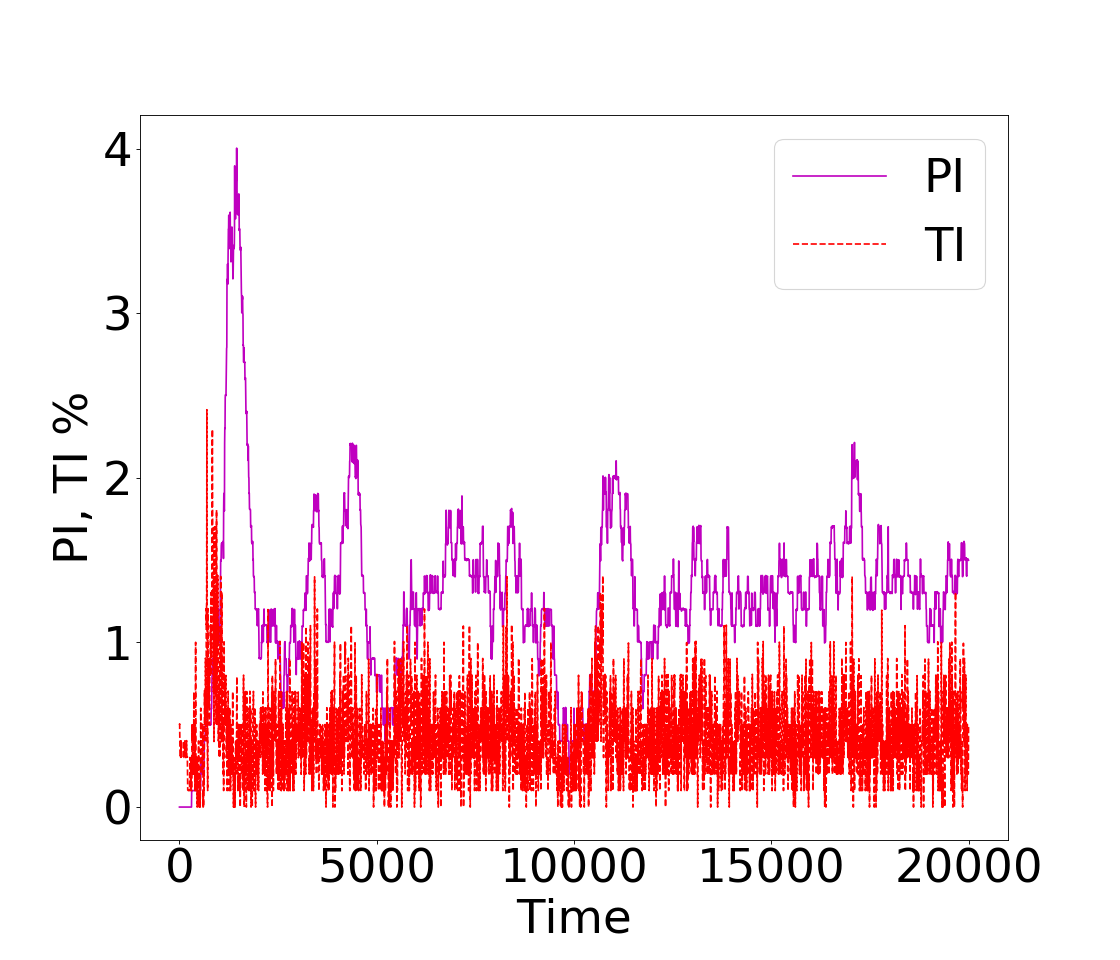}
		\caption{Infectious states' evolution.}
		\label{fig:1farmIP_wellP0}
	\end{subfigure}

	\begin{subfigure}[b]{0.45\textwidth}
	\includegraphics[width=\textwidth]{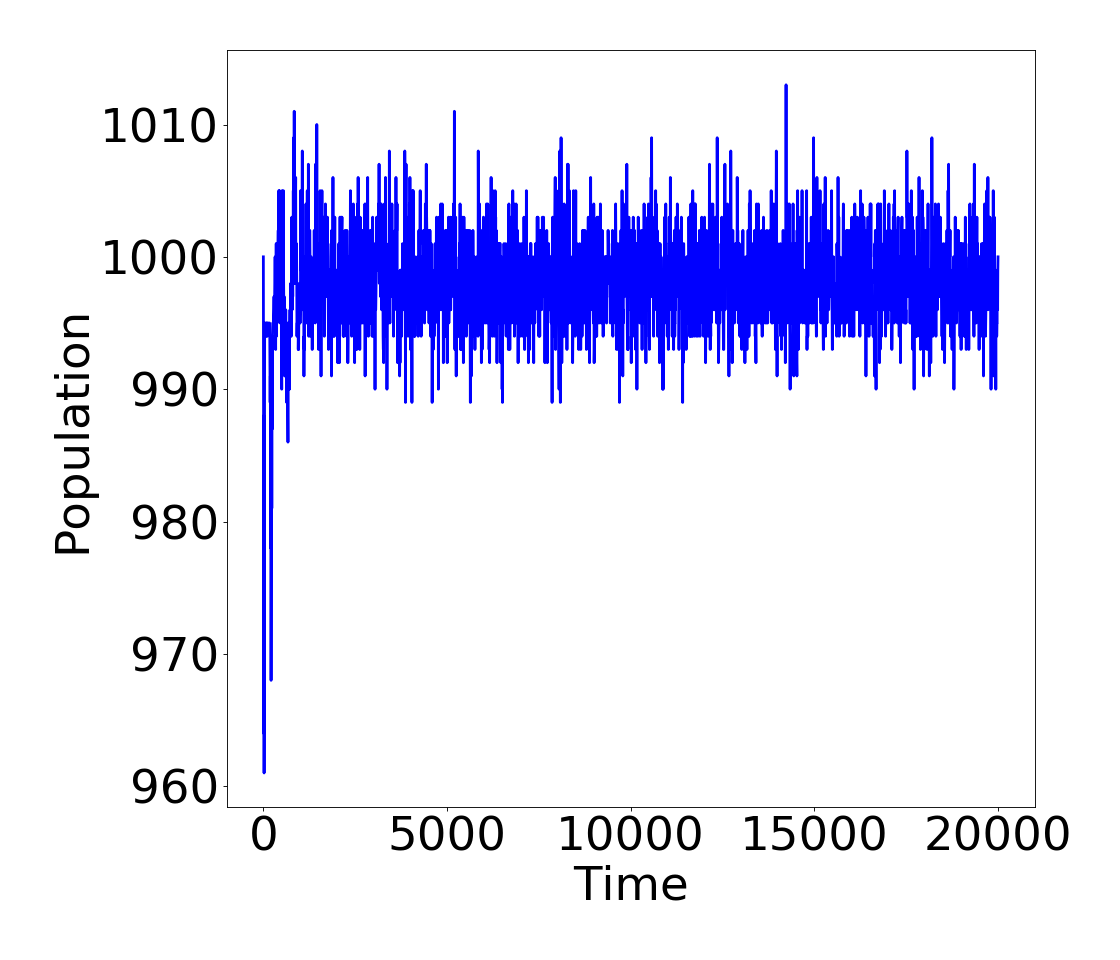}
	\caption{Demographic evolution.}
	\label{fig:1farm_demog}
	\end{subfigure}
	\caption{All the S, I, R and P states in a single graph \ref{fig:1farmSIRP_wellP0} for the single farm with no PI inputs and only the infectious I and P in \ref{fig:1farmIP_wellP0}. Moreover, we see in \ref{fig:1farm_demog} the evolution of the same farm's population for no intervention strategy with animal inputs and outputs (source and drain farm) to satisfy its demand that its population be constant.}\label{fig:1farm_wellP0}
\end{figure}

We develop some intuition on a minimal example of how the system should behave in terms of infectious states and population in figures \ref{fig:1farmSIRP_wellP0}, \ref{fig:1farmIP_wellP0} and \ref{fig:1farm_demog} respectively. The simulation runs for the same settings as in the main text without any intervention strategy, as calibrated for the case of Germany in \cite{BAS18} and with no PI animals originating from the source farm. The population of the single farm is set to 1,000 animals to achieve well-mixed conditions for the epidemic dynamics and to diminish finite-size effects. Since there is an inherent supply and demand mechanism to equalize the imbalance caused by demographic factors in the system, the minimum working setup requires the source and the drain farms as well. We notice therefore that the population fluctuates around the initial value of 1,000.

Moreover, the peak of infection is much more pronounced when only a single farm is considered as can be seen in figure \ref{fig:1farmIP_wellP0}. This effect is typical of SIR dynamics and depends on the involved infectious and recovery rates \cite{KEE08}. The spatial structure may then shift the outbreak peak in value and time depending on the exact heterogeneities (farm size distribution) and connectivity (supply and demand) of the farm-nodes involved in the simulation.

\subsubsection*{Effect of PIs from the Source Farm}

For the single farm with 2\% probability of introduction of PI animals through the source farm, the demographic time-series is virtually identical to the one for no PI animals entering the system through the source farm. The settings remain always as those in the main text without any intervention strategy.

\begin{figure}
	\centering
	\begin{subfigure}[b]{0.45\textwidth}
		\includegraphics[width=\textwidth]{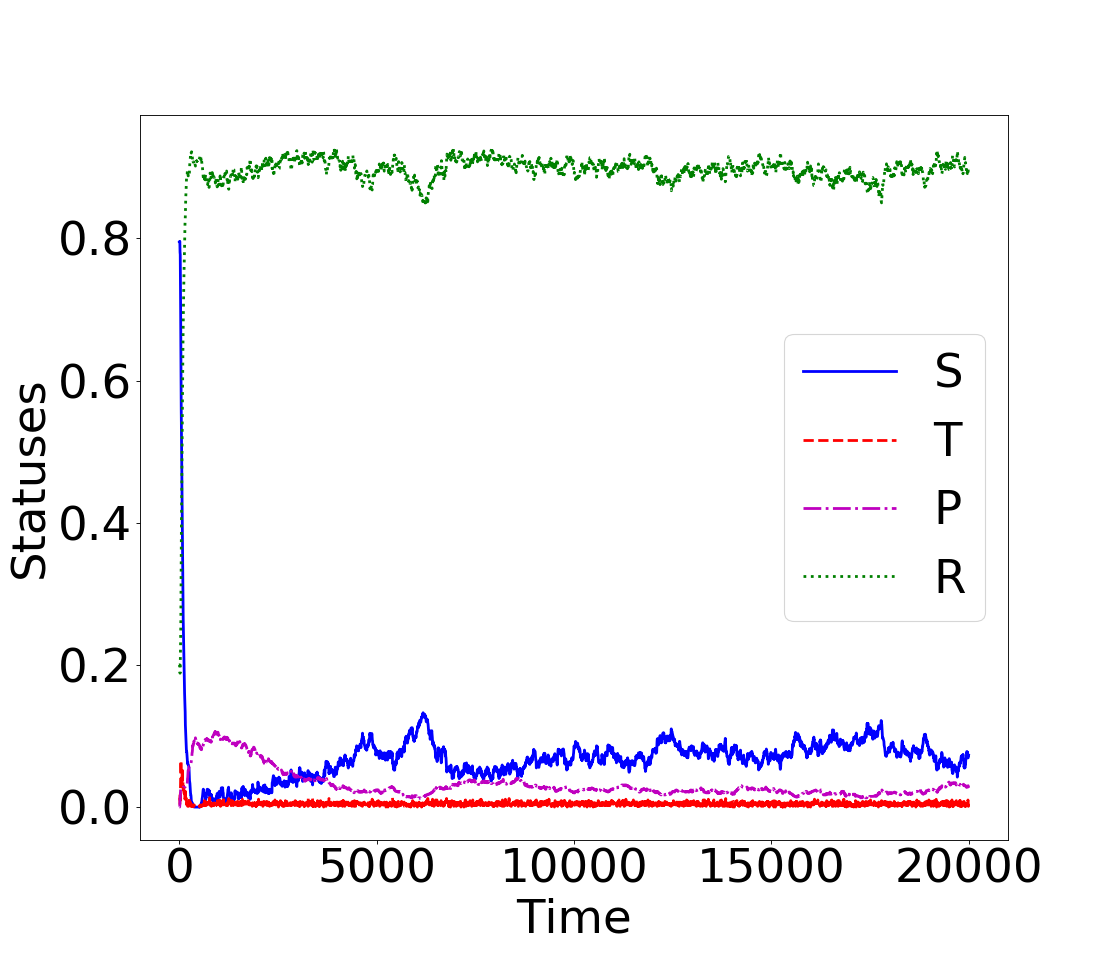}
		\caption{All states' evolution.}
		\label{fig:1farmSIRP_well_PI2}
	\end{subfigure}
	~
	\begin{subfigure}[b]{0.45\textwidth}
		\includegraphics[width=\textwidth]{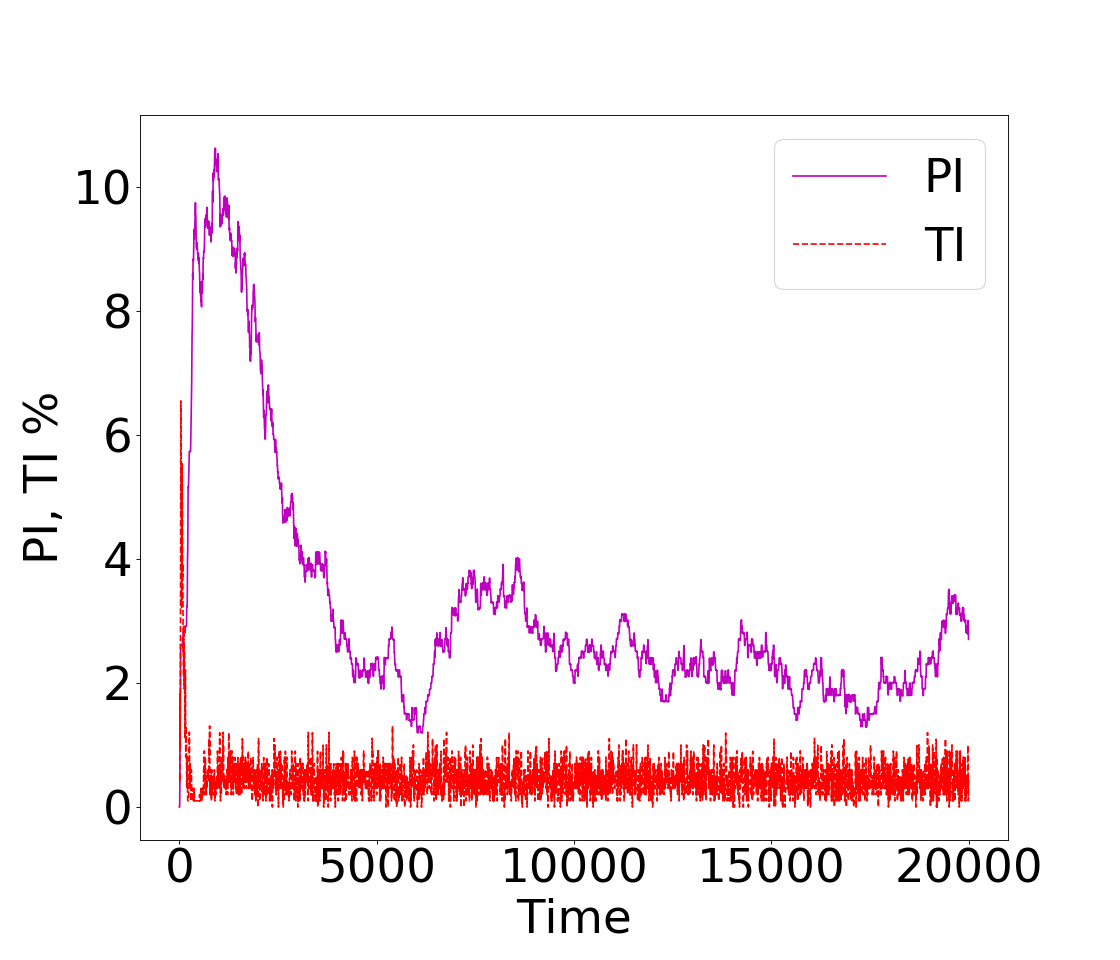}
		\caption{Infectious states' evolution.}
		\label{fig:1farmIP_well_PI2}
	\end{subfigure}
	\caption{The respective evolution plots of figure \ref{fig:1farm_wellP0} but for a probability of a PI introduction from the source farm of 2\%.}\label{fig:1farm_well_PI2}
\end{figure}

What stands out in figures \ref{fig:1farmSIRP_well_PI2} and \ref{fig:1farmIP_well_PI2} in respect to their counterparts \ref{fig:1farmSIRP_wellP0} and \ref{fig:1farmIP_wellP0} (the demographics are statistically identical in both cases and are therefore skipped in figure \ref{fig:1farm_well_PI2}) of no PI animals' introduction from the source farm is the epidemic outbreak peak value and the effect this has on the immune, R class of animals. The PI peak is over twice times higher than in the former case and therefore the R class of animals remains consistently at high levels, without much fluctuation. This is a consequence of the increased secondary, transient transmissions induced to susceptible animals from the case of introduced PI animals in the system. These behaviors are both intuitively expected as the number of secondary, transient infections should be mainly proportional to the number of PI animals in the system, them being the main contributor of the epidemic spread for both the well-mixed (local) and spatially distributed (network) dynamics.

\subsection*{Sensitivity Analysis}

In the following we make a sensitivity analysis on four factors dominating over the epidemic, testing and vaccination dynamics, as well as on the stochastic fluctuations of the simulations overall.

\begin{enumerate}
	
	\item Transmission Rates
	
	The catalytic factor affecting the final state of the PI animals is shown to be the PI transmission coefficient $\beta_{PI}$ as demonstrated in table \ref{tab:beta_sens}. This result is somewhat contrary to that of \cite{EZA07} mainly because the between-herd (animal group in the authors' case) dynamics were not considered in our simulation as opposed to that work. To be precise, despite the form of the infectious dynamics being nearly identical (see \cite{BAS18}), the authors of \cite{EZA07} considered only a five animal herd group, therefore introducing pronounced finite-size effects in their results. Furthermore, in the simulation of our work, due to the effect of vertical transmissions, i.e. PI animals latently appearing in the system after a pregnant cow's infection, the effect of the PI transmission coefficient $\beta_{\text{PI}}$ dominates over $\beta_{\text{TI}}$. In addition, the sensitivity analysis of \cite{EZA07} on the PI animals' death rates is not directly comparable to our work, as we presume a uniform random distribution for that purpose.
	
	\begin{table}
		\begin{center}
			\begin{tabular}{|c||c|c|} \hline
				PI\textsubscript{final} No & $\beta_{\text{TI}}$ & $\beta_{\text{PI}}$  \\ \hline
				11 & 0.01 & 0.1 \\
				210 & 0.01 & 0.5 \\
				278 & 0.01 & 0.8 \\
				12 & 0.03 & 0.1 \\
				305 & 0.03 & 0.5 \\
				12 & 0.05 & 0.1 \\
				317 & 0.05 & 0.8 \\
				21 & 0.1 & 0.1 \\ \hline
			\end{tabular}
			\caption[Final number of persistently infected animals as a function of $\beta_{\text{TI}}$ and $\beta_{\text{PI}}$.]{Final number of persistently infected animals as a function of $\beta_{\text{TI}}$ and $\beta_{\text{PI}}$.}\label{tab:beta_sens}
		\end{center}
	\end{table}
	
	\item Ear Tag and Retesting Laps
	
	For the tests' effect on the PI prevalence their sensitivity was examined by varying the test's accuracy probability as seen in figure \ref{fig:ear_tag_sens}. Not surprisingly, as the sensitivity probability tends to unity the PI prevalence tends to eradication as all the infected animals are identified and subsequently sent to the drain farm (slaughterhouse).
	
	\begin{figure}
		\centering
		\includegraphics[width=\linewidth]{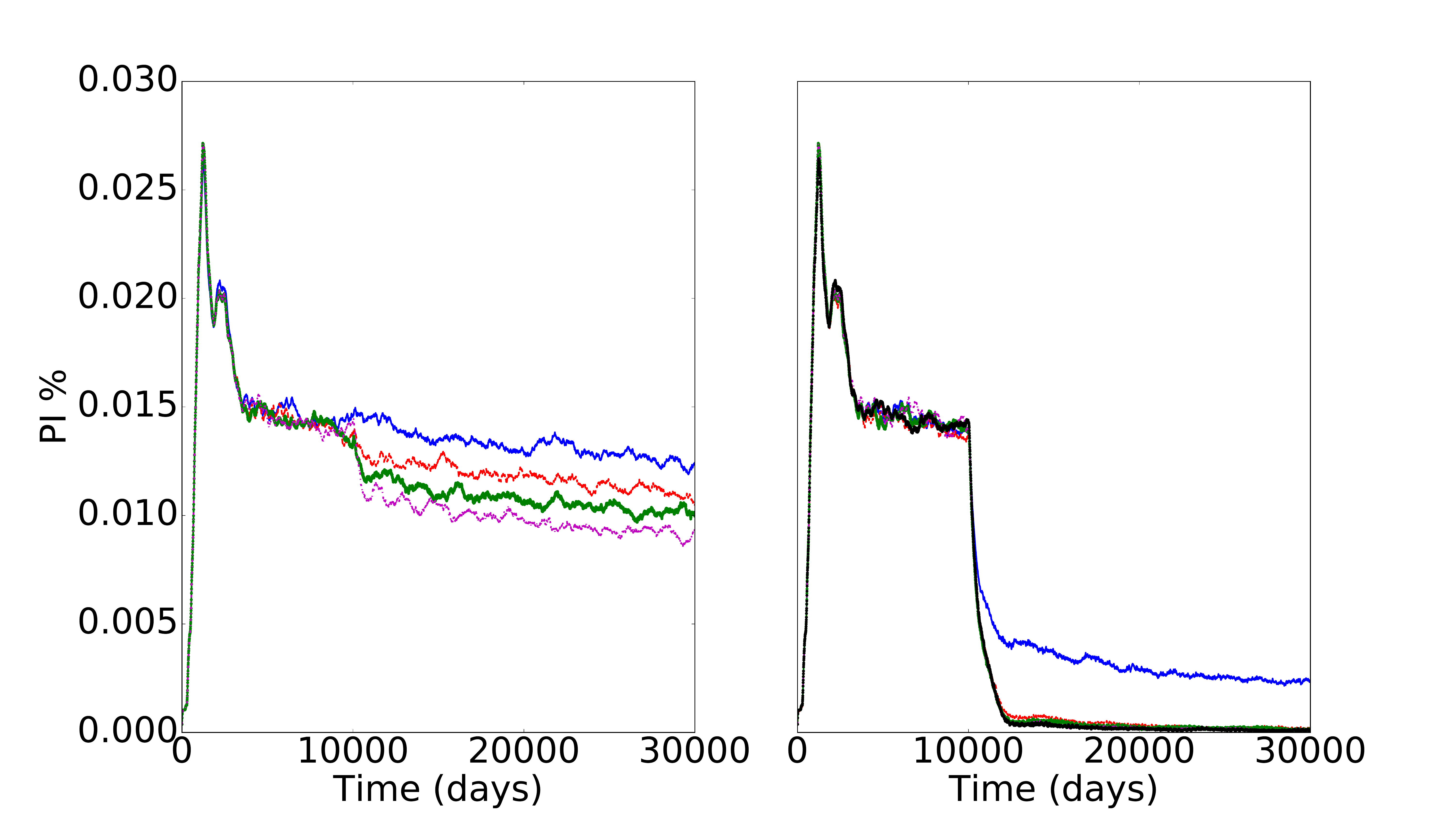}
		\caption[Scenario 3 (new regulation) for different test accuracy probabilities ranging from 0 to 1.]{Scenario 3 (reference the simulation plan from the main text) for different test accuracy probabilities ranging from 0 to 1, as outlined in table \ref{tab:ear_tag}. The effect of the ear tag test protocol is enforced from day 10,000 onwards.}
		\label{fig:ear_tag_sens}
	\end{figure}
	
	\begin{table}
		\begin{center}
			\begin{tabular}{|c|c|} \hline
				Probability & Style (Position) \\ \hline
				0 & Solid, blue (left)\\
				0.1 & Dashed, red (left) \\
				0.2 & Dotted-dashed, green (left) \\
				0.3 & Dashed, dotted, magenta (left) \\
				0.8 & Solid, blue (right) \\ 
				0.98 & Dashed, red (right) \\
				0.99 & Dotted-dashed, green (right) \\
				0.998 & Dashed, dotted, magenta (right) \\ 
				1 & Dotted, black (right)  \\\hline
			\end{tabular}
			\caption[Probabilities of the test's sensitivity for scenario 3 (new regulation).]{Probabilities of the test's sensitivity for scenario 3 (reference the simulation plan from the main text).}\label{tab:ear_tag}
		\end{center}
	\end{table}
	
	\item Vaccination Success Probability
	
	For the vaccination's effect on the PI prevalence the vaccination's working probability was varied from zero to one as demonstrated in figure \ref{fig:vac_work_prob}. As with the tests' sensitivity success probability, the increase of certainty for the vaccine's working probability leads to eventual extinction of the PI population. However, the vaccination's effect on the PI population is slower than that of the tests' sensitivity success and requires near certainty values to lead to extinction. This could be attributed to the fact that indiscriminate vaccination of animals regardless of their infectious status reaches its intended targets slower than tests of all animals, which aim to identify and remove the infected ones.
	
	\begin{figure}
		\centering
		\includegraphics[width=\linewidth]{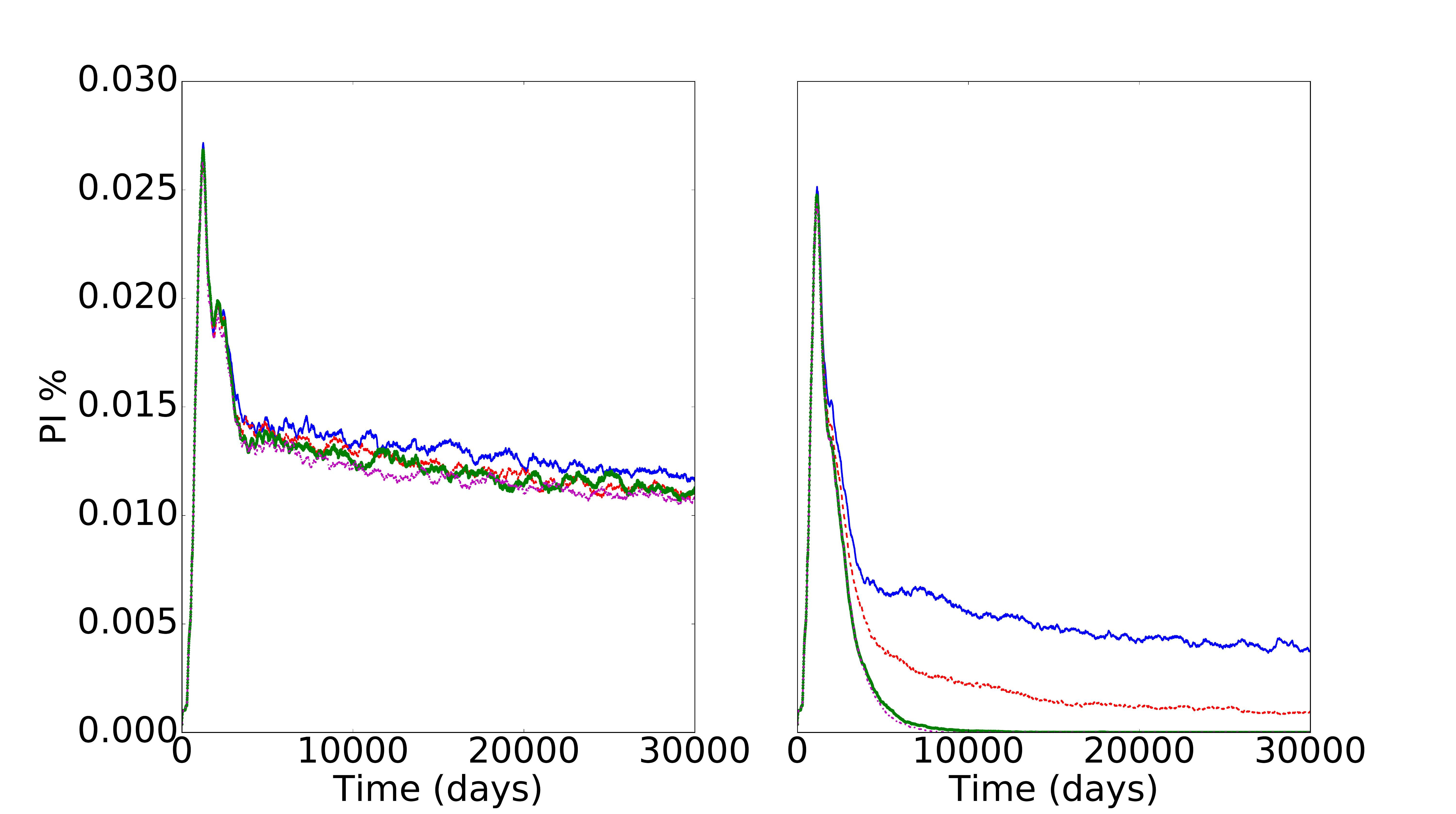}
		\caption[Strategy 8 (vaccination) for different vaccination working probabilities ranging from 0 to 1.]{Strategy 8 (reference the simulation plan from the main text) for different vaccination working probabilities ranging from 0 to 1 as outlined in table \ref{tab:work_vac}.}
		\label{fig:vac_work_prob}
	\end{figure}
	
	\begin{table}
		\begin{center}
			\begin{tabular}{|c|c|} \hline
				Probability & Style (Position) \\ \hline
				0 & Solid, blue (left)\\
				0.1 & Dashed, red (left) \\
				0.2 & Dotted-dashed, green (left) \\
				0.3 & Dashed, dotted, magenta (left) \\
				0.8 & Solid, blue (right)  \\
				0.9 & Red, dashed (right) \\ 
				0.985 (default) & Dotted-dashed, green (right) \\
				1 & Dashed-dotted, magenta (right) \\ \hline
			\end{tabular}
			\caption[Vaccination working probabilities for the sensitivity of strategy 8 (vaccination).]{Vaccination working probabilities for the sensitivity of strategy 8 (reference the simulation plan from the main text).}\label{tab:work_vac}
		\end{center}
	\end{table}
	
	\item Variance
	
	For completeness and reliability of the stochastic nature of the simulation a sensitivity analysis was performed on the PI prevalence percentage as a function of the seeds of the random number generator as displayed in figure \ref{fig:variance}. The differences proved to be of the order of 0.001\% implying that the results are robust to stochastic fluctuations.
	
	\begin{figure}
		\centering
		\includegraphics[width=\linewidth]{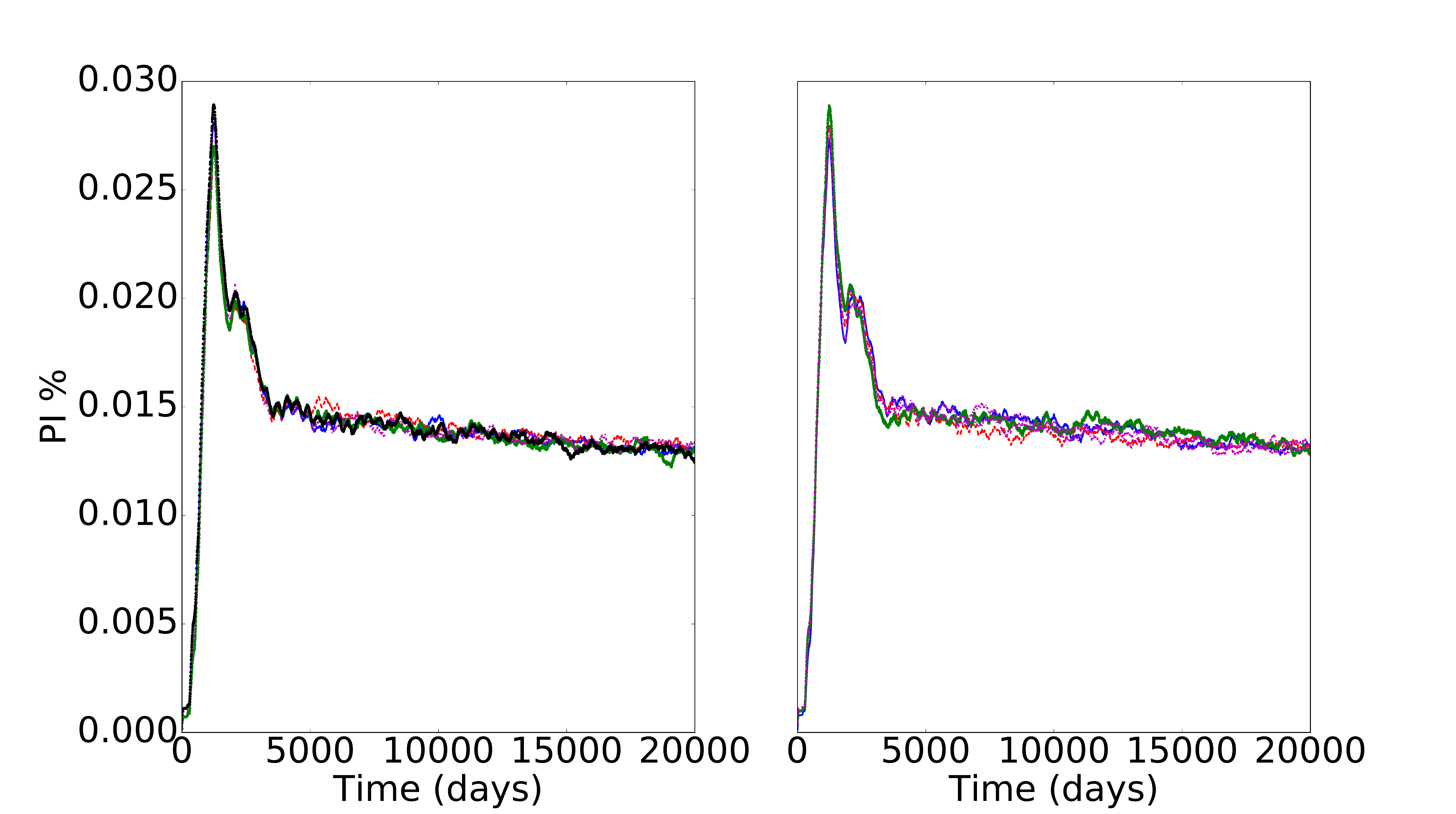}
		\caption{Scenario 1 (reference the simulation plan from the main text) for different seeds as outlined in table \ref{tab:var_seeds}.}
		\label{fig:variance}
	\end{figure}
	
	\begin{table}
		\begin{center}
			\begin{tabular}{|c|c|} \hline
				Seed No & Style (Position) \\ \hline
				2,333,600,960 & Solid, blue (left)\\
				2,333,600,970 & Dashed, red (left) \\
				2,333,601,960 & Dotted-dashed, green (left) \\
				2,333,620,963 & Dashed, dotted, magenta (left) \\
				2,333,710,962 & Dotted, black (left)  \\
				2,333,970,967 & Solid, blue (right) \\ 
				2,333,650,932 & Dashed, red (right) \\
				4,333,600,860 & Dotted-dashed, green (right) \\
				3,323,601,969 & Dashed, dotted, magenta (right) \\ 
				3,323,601,970 & Dotted, black (right) \\ \hline
			\end{tabular}
			\caption{Seeds for the variance of scenario 1 (baseline).}\label{tab:var_seeds}
		\end{center}
	\end{table}
	
\end{enumerate}

\end{document}